\documentclass[journal]{IEEEtran}
\usepackage{amsmath}
\usepackage{lineno,hyperref}
\modulolinenumbers[5]
\usepackage{amssymb}
\usepackage{booktabs}
\usepackage{graphicx}
\usepackage{bm}
\usepackage{cite}
\usepackage{multirow}
\usepackage{booktabs}
\usepackage[table,xcdraw]{xcolor}
\usepackage{pifont}
\usepackage{bbm}
\usepackage{float}
\begin{document}

	\title{ContrasInver: Ultra-Sparse Label Semi-supervised Regression for  Multi-dimensional Seismic Inversion}
	\author{Yimin Dou, Kewen Li, Wenjun Lv, Timing Li, Zhifeng Xu
		\thanks{\textit{Corresponding author: Kewen Li}}
		\thanks{Yimin Dou, Kewen Li, Zhifeng Xu College of computer science and technology, China University of Petroleum (East China) Qingdao, China.}
		\thanks{Wenjun Lv, Department of Automation, University of Science and Technology of China, China.}
		\thanks{Timing Li, College of Intelligence and Computing, Tianjin University Tianjin, China.}
		%\thanks{Hongjie Duan, Weidong Cao, Shengli Oilfield Company, SINOPEC Dongying, China.}
		\thanks{This work was supported by grntsa from the National Natural Science Foundation of China (Major Program, No.51991365), and the Natural Science Foundation of Shandong Province, China (ZR2021MF082).}
	}
	\maketitle
	
\begin{abstract}	
The automated interpretation and inversion of seismic data have advanced significantly with the development of Deep Learning (DL) methods.
However, these methods often require numerous costly well logs, limiting their application only to mature or synthetic data. This paper presents ContrasInver, a method that achieves seismic inversion using as few as two or three well logs, significantly reducing current requirements. 
In ContrasInver, we propose three key innovations to address the challenges of applying semi-supervised learning to regression tasks with ultra-sparse labels.
The Multi-dimensional Sample Generation (MSG) technique pioneers a paradigm for sample generation in multi-dimensional inversion. It produces a large number of diverse samples from a single well, while establishing lateral continuity in seismic data. MSG yields substantial improvements over current techniques, even without the use of semi-supervised learning.
The Region-Growing Training (RGT) strategy leverages the inherent continuity of seismic data, effectively propagating accuracy from closer to more distant regions based on the proximity of well logs. 
The Impedance Vectorization Projection (IVP) vectorizes impedance values and performs semi-supervised learning in a compressed space. We demonstrated that the Jacobian matrix derived from this space can filter out some outlier components in pseudo-label vectors, thereby solving the value confusion issue in semi-supervised regression learning.
In the experiments, ContrasInver achieved state-of-the-art performance in the synthetic data SEAM I. In the field data with two or three well logs, only the methods based on the components proposed in this paper were able to achieve reasonable results. It's the first data-driven approach yielding reliable results on the Netherlands F3 and Delft, using only three and two well logs respectively.

\end{abstract}

\section{Introduction}
Oil is called the blood of the industry, recent studies have shown that learning theories have been very successful in hydrocarbon exploration\cite{iqbal2022deepseg,dou2021attention,dou2022md,chang2021seglog,chang2021active}. Impedance estimation is a critical step in characterizing hydrocarbon reservoirs from exploration data (seismic data)\cite{latimer2000interpreter}. Current Deep Learning (DL) methods require geologically similar pre-trained models or dozens to even more logs to achieve promising results\cite{das2019convolutional,wu2021deep,meng2021seismic,wu2022seismic,wang2020well,wang2021physics,wang2022seismic,xie2022seismic}. 
In fact, most new and special oil and gas fields cannot meet this requirement. Furthermore, logging is very expensive, often costing millions to tens millions dollars for a single logged well. Therefore, the use of fewer well logs to accurately estimate impedance is of high engineering and economic value to exploration efforts.

\begin{figure}[!t]
	\centering
	\includegraphics[scale=0.15]{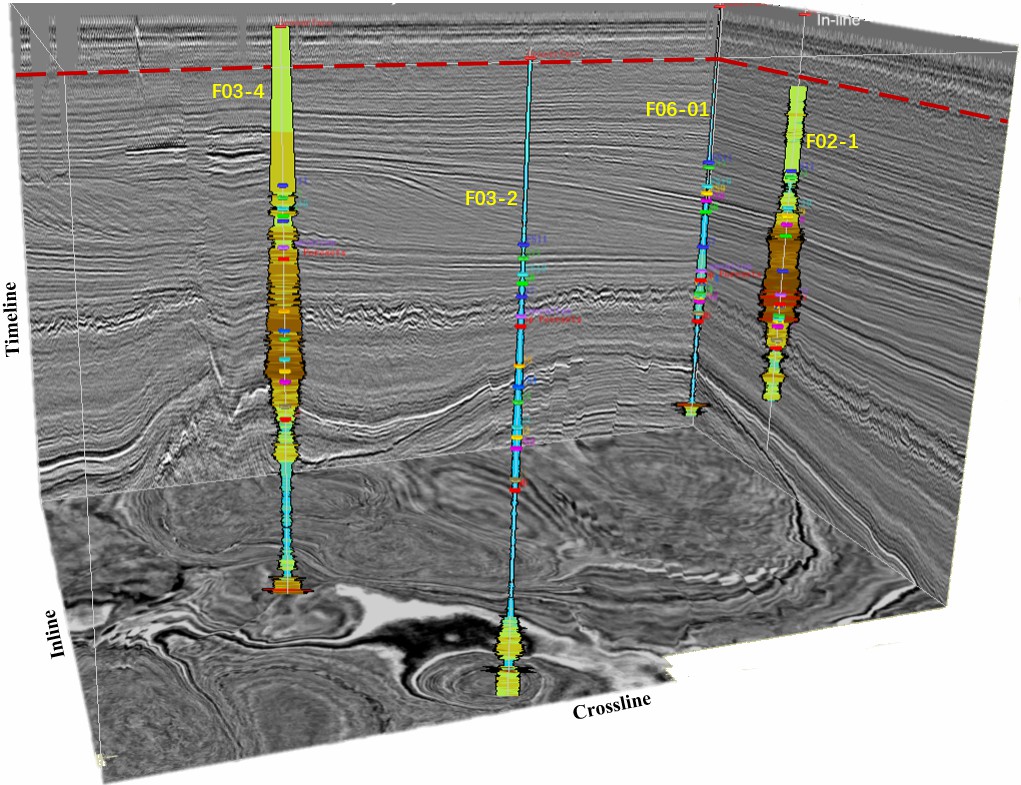}
	\centering\caption{This paper examines the data-driven impedance inversion task, utilizing the Netherlands F3 as an illustrative example. The figure demonstrates a complete seismic volume along with four well logs.
		The seismic data is denoted as $\textbf{D} \in \mathbb{R}^{t\times l_\text{i} \times l_\text{x}}$, where $t$ is the seismic trace length, $l_\text{i}$ is the inline length, and $l_\text{x}$ is the crossline length. The well log data (labels) are represented as $y_i \in \mathbb{R}^{\phi_i t \times1 \times 1}$, where $y_i$ is a vector of continuous values, $\phi_i$ is the ratio of well log to seismic traces. Each $y_i$ corresponds to a seismic trace $x_i$ within the seismic volume.
		As depicted in the figure, impedance inversion is a regression task where a 3D seismic volume learns from 1D continuous value labels.
	}

	\label{seismic-log}
\end{figure}

Theory-driven inversion is the common means of impedance estimation in traditional hydrocarbon exploration, which includes sparse-based\cite{wang2016seismic,sui2019nonstationary}, model-based\cite{veeken2004seismic,wu2017structure} and other methods\cite{guo2019hybrid,bosch2010seismic}. These approaches are more dependent on the initial model and the hyperparameter settings, and here we focus on the learning task of data-driven impedance estimation. Through Fig. \ref{seismic-log}, we simplify the inversion problem into a task that data scientists can easily understand, that is, the regression learning task between 3D data and 1D labels. In the case of few labels, this task is also very attractive and challenging in the field of machine learning.

\subsection{Recent DL-based Inversion methods}
In the beginning, researchers tried to find mapping relationships between seismic traces that matched the logging dimension. Hampson demonstrated that impedance can be predicted by combining multiple seismic traces and corresponding attributes using a simple neural network\cite{hampson2001use}. Recently an increasing amount of work has starting use DL to estimate impedance\cite{das2019convolutional,wu2021deep,meng2021seismic,wu2022seismic,wang2020well,wang2021physics,wang2022seismic,xie2022seismic}. 
However, it is well known that DL relies on significant amounts of data, so some of these works use semi-supervised methods. Wu et al. used a semi-supervised method based on adversarial learning\cite{wu2021semi,meng2020semi}, a prototype of which was first proposed by Wei\cite{hung2018adversarial}, who trained the discriminator to distinguish between confidence maps from labeled and unlabeled data predictions.
The method depends on having enough labels to ensure the stability of the Generative Adversarial Networks (GAN), so the inversion of the SEAM I still requires 34 logging labels, and its predictions have significant discontinuities in the horizontal direction, the same drawback is also reflected in other semi-supervised impedance inversion methods \cite{ge2022semi,song2022semi,di2020semi}.
The discontinuity in the horizontal direction is due to the fact that these methods try to match the dimensionality of the logs by downscaling the 3D or 2D seismic data, which is avoided by the multidimensional inversion proposed by Wu et al \cite{wu2021deep}, the idea of this method originates from medical image segmentation\cite{cciccek20163d}, where the model is trained using labels of the same dimensionality as the seismic and the weights of the unlabeled regions are set to zero. The method achieved a significant performance improvement with the use of 40 logs in SEAM.

In conclusion, the inversion task presents a completely new context for machine learning or deep learning. Data-driven inversion has not yet received widespread attention, and there are many important challenges in this field that have not been deeply researched or mentioned.

\subsection{Motivation and Contributions}
\subsubsection{The challenge of data driven inversion}
The current data-driven methods for seismic inversion suffer from several challenges.

\textit{The scarcity of labels:} Well logging demand, high for both traditional and recent methods, often limits techniques to mature or synthetic data. Applying these to new or special oil fields is hard. With only three or four logs, classic F3 and Delft surveys have had no successful inversions.

\textit{The inconsistency of geological structures:}  Many 1D methods predict seismic traces individually, causing inconsistency in geological structures and lateral discontinuity\cite{das2019convolutional,wu2022seismic}. Wu's 2D method\cite{wu2021deep}, despite its strict sampling and well log support needs, only ensures one-direction continuity for 3D or higher-dimensional data.

\subsubsection{The challenge of semi-supervised inversion}
In response to the aforementioned challenges, one plausible approach is semi-supervised learning. However, currently, there is no existing framework that can directly address these challenges and be applied to seismic data inversion. To apply the most advanced machine learning methods to seismic data inversion, the following issues need to be resolved.

\textit{Ultra-sparse labeling in geophysical scenarios:}  Dense semi-supervised methods require around 3.33\% (Cityscapes) or 6.25\% (VOC) labeled samples for effective pre-training\cite{chen2021semi,wang2022semi}. However, geophysical scenarios often present less than 0.01\% labeled well log data. The key challenge is maximizing the sparse well log data use for sample construction for semi-supervised learning. How samples are constructed impacts the learning process and framework design.

\textit{Insufficient semi-supervised methods for sparse regression labels:} Most semi-supervised research focuses on classification or segmentation, with a gap in regression. Common methods include pseudo-labeling\cite{lee2013pseudo}, consistency\cite{sajjadi2016regularization,laine2016temporal}, adversarial learning (GAN)\cite{hung2018adversarial}, and contrastive\cite{zhong2021pixel}, typically studied for discrete and complete labeling. However, they're not often applied for regression tasks with continuous, sparse labels, where their direct use may yield subpar results due to ambiguous labels or features.

%\textbf{No methods can be deployed straightforwardly or with tweaks:} 
\textit{Lack of precedent tasks as benchmark:}
Multi-dimensional semi-supervised impedance inversion provides a fresh context for machine learning, but remains unexplored. Existing semi-supervised methods can't be easily applied or tweaked for this task. Creating a learning framework that can glean impedance information from ultra-sparse 1D well logging in 3D seismic data is unique and challenging. This approach, due to its reliance on well log-derived labels, could have wide application in geophysical exploration tasks like impedance inversion, velocity, density, and lithology assessment.

\subsubsection{Contributions}

We've developed ContrasInver, a semi-supervised inversion framework requiring only two or three well logs for reliable inversion—about a tenth of existing methods. Applicable to any dimensional data without strict preprocessing or low-frequency impedance constraints, ContrasInver follows two stages: pre-training and semi-supervised training. It utilizes a classic dual-network structure (Mean Teacher \cite{wang2022dual}), and supported by our three main innovations, addressing the outlined challenges.

\textit{Multi-dimensional Sample Generation: }
In pre-training, we introduced a novel Multi-dimensional Sample Generation (MSG) for seismic inversion. MSG enables tens of thousands of training samples from each well log without data augmentation, reducing overfitting risk in sparse well log scenarios. Even without semi-supervised process, data-driven inversion with MSG significantly outperforms existing methods.

\textit{Region-Growing Training Strategy:}
Given seismic data's continuity, the accuracy of impedance inversion in a pre-trained model is inversely proportional to its distance from well logs—the closer to well log coordinates, the higher the accuracy. Hence, we introduce a Region-Growing Training (RGT) strategy. Centered on well logs, inner rings supervise outer rings through overlapping areas, facilitating semi-supervision. The outer rings thereby learn more accurate information, spreading globally from near to far. We've also devised an efficient method to calculate overlapping areas of random samples, enhancing training speed and allowing real-time performance even for high-resolution 3D cubes.

\textit{Impedance Vectorization Projection: }
To address the value confusion issue (inability to evaluate the quality of pseudo-labels or features) in semi-supervised learning for regression tasks, we introduce Impedance Vectorization Projection (IVP). IVP transforms the model's output into impedance vectors in a base vector-defined space, projecting these vectors to a lower-dimensional space. A contrastive loss in this compressed space minimizes feature distance discrepancies between two networks, with vector directions constrained by well logging. 
We demonstrated that the Jacobian matrix derived from the projection space can filter out some gradients caused by outliers. Our semi-supervised ContrasInver framework, particularly effective in regression tasks, leverages vectorized projection to mitigate learning value confusion, offering an alternative to direct pseudo-label or EMA model-generated feature learning. This can extend to other semi-supervised regression tasks like counting, pose estimation, target regression, etc.

Our unsupervised learning process is based on a contrastive proxy task (RGT+IVP), making it a form of contrastive semi-supervised learning \cite{zhong2021pixel,yang2022class,lee2022contrastive,alonso2021semi}. Therefore, this framework is referred to as ContrasInver.

\section{Related Works}
Multi-dimensional impedance inversion presents a completely new context for the machine learning community. In the introduction, we have analyzed some data-driven methods that have already been applied in this field, and these methods have various shortcomings. In this chapter, we analyze some of the more advanced semi-supervised or self-supervised methods, attempting to identify which methods can be directly applied or potentially applied in this field.

\subsection{Semi-supervised Learning}

\subsubsection{Pseudo label} 
%Semi-supervision wants to learn from unlabeled data, and creating pseudo-labels is a common approach\cite{lee2013pseudo}. Most methods choose high confidence predictions as pseudo-labels\cite{chen2021semi,guo2022class,yang2022st++,zou2020pseudoseg,zuo2021self}, which can also lead to most unlabeled data being discarded due to their unreliability, and some recent work has started to focus on lower confidence prediction values\cite{wang2022semi}.
%Existing methods for creating pseudo-labels are suitable for classification and segmentation tasks that employ discrete labels. The target of the regression task is the continuous value, it is not possible to create pseudo-labels based on confidence, and blindly using unfiltered pseudo-labels can degrade model performance\cite{arazo2020pseudo}. Therefore it is difficult to use such methods in impedance estimation tasks if a reasonable way cannot be found to evaluate the quality of the network output. 

Semi-supervision relies on creating pseudo-labels, typically through high confidence predictions\cite{lee2013pseudo,guo2022class,yang2022st++}. However, this approach often discards unreliable unlabeled data, and recent studies have explored using lower confidence values\cite{wang2022semi}. Yet, these methods are designed for discrete labels in classification and segmentation tasks, making it challenging to apply them to regression tasks that involve continuous values. Blindly using unfiltered pseudo-labels can degrade model performance\cite{arazo2020pseudo}. Thus, finding a reasonable way to evaluate network output quality is crucial for impedance estimation tasks.

\subsubsection{Consistency regularization} 
Consistency regularization ensures that a model produces consistent outputs with different data augmentations. Approaches like \cite{sajjadi2016regularization,laine2016temporal} use mean square error for consistency measurement, while Mean Teacher was introduced by Tarvainen et al.\cite{tarvainen2017mean}. VAT replaces traditional transformations with adversarial transformations\cite{miyato2018virtual}, and MixMatch improves stability by averaging predictions from multiple augmented samples\cite{berthelot2019mixmatch}. UDA, ReMixMatch, and FixMatch adopt cross-entropy loss and robust augmentations\cite{xie2020unsupervised,berthelot2019remixmatch,sohn2020fixmatch}. CCT incorporates data perturbation based on semantic segmentation clustering\cite{ouali2020semi}. The DM2T-Net encourages multiple predictions at different CNN layers for consistency while computing multi-scale loss\cite{wang2022dual}.

Consistency regularization in semi-supervised learning does not require output discretization. Instead, it focuses on learning the network's output probabilities directly. In regression tasks like impedance inversion, continuous network outputs serve as the final results without discretization. However, predictions from weakly augmented samples may lack accuracy, introducing value confusion that can hinder the learning process.

\subsubsection{Generative Adversarial Network (GAN)} 
GAN has shown promise in semi-supervised learning for regression tasks, particularly in image matting\cite{ke2020guided}. Scholars have applied GAN to semi-supervised impedance inversion\cite{wu2021semi,meng2020semi} and natural/medical image segmentation\cite{hung2018adversarial,mittal2019semi,perone2018deep,feng2022dmt,feng2020semi,zhai2022ass,xie2019retinopathy}. Hung et al. introduced discriminators to differentiate between confidence maps of labeled and unlabeled data\cite{hung2018adversarial}. Mittal et al. employed a two-branch approach with GAN for low entropy predictions and false positive elimination\cite{mittal2019semi}. Similar approaches were followed by Feng et al.\cite{feng2022dmt,feng2020semi}. Pseudo-labels in these methods are generated by GAN, not based on confidence. SDA-GAN by Dong et al. used a domain alignment module to reduce distribution gaps\cite{dong2022semi}. AALLI by Chang and Lv focused on domain adaptation for well logging lithology identification\cite{chang2021active}. However, stable GAN training still requires sufficient labeled data\cite{hung2018adversarial}. The main limitation is that GAN training requires complete labeling for each sample, limiting its use to 1D network training and resulting in lateral discontinuity.

\subsubsection{Contrastive semi-supervised learning} 
%Contrastive semi-supervised underscores the importance of the class distinguishability of pixel-level features\cite{zhong2021pixel,yang2022class,lee2022contrastive,alonso2021semi}, distinguishing it from pseudo-labeling and consistency methods. It places a greater emphasis on maintaining consistency across feature domains and employs metric losses (contrastive losses). Consequently, it can be considered as a fusion of consistency regularization and contrastive learning \cite{lee2022contrastive}. Correspondingly, it also faces a similar challenge as consistency regularization, which involves the need to identify feature vectors corresponding to high-quality labels. Currently, the predominant methods utilized still rely on confidence-based approaches to address this matter\cite{alonso2021semi,lee2022contrastive}.
Contrastive semi-supervised learning focuses on the distinguishability of pixel-level features for better class separation\cite{zhong2021pixel,yang2022class,lee2022contrastive,alonso2021semi}, its unsupervised process is conducted through a contrastive proxy task. It differs from pseudo-labeling and consistency methods by emphasizing consistency across feature domains and using contrastive losses. It can be seen as a fusion of consistency regularization and contrastive learning\cite{lee2022contrastive}. However, it faces a similar challenge as consistency regularization in identifying feature vectors corresponding to high-quality labels. Currently, confidence-based approaches are predominantly used to address this challenge\cite{alonso2021semi,lee2022contrastive}.

\subsection{Contrastive Learning}\label{constrabackg}
Self-supervised and unsupervised representation learning have made significant progress by leveraging contrastive learning. These methods utilize large-scale datasets to train pre-trained models, enabling the transfer of information to downstream tasks \cite{chen2020simple,he2020momentum,grill2020bootstrap}. Recently, there has been extensive exploration of pixel-level contrastive learning, which is particularly well-suited for tasks such as object detection and semantic segmentation \cite{xie2021propagate,wang2021dense,liu2022contrastive}. In this paper, we aim to investigate the potential benefits of semi-supervision for models, moving beyond the scope of solely building pre-trained models. Therefore, this work specifically focuses on the fusion of contrastive thinking with semi-supervised learning to address seismic inversion problems.

%\subsection{Discussion on Related Works}
%As emphasized in the introduction, multidimensional inversion introduces a new task to the machine learning community. In the related works we have discussed, traditional semi-supervised tasks in image analysis differ significantly from seismic image inversion. Existing methods cannot be directly applied to this task, leading to three major challenges:
%
%\begin{itemize}
%	\item[$\bullet$] Ultra-sparse labeling in geophysical scenarios.
%	\item[$\bullet$] Insufficient semi-supervised methods for sparse regression labels.
%	\item[$\bullet$] Lack of precedent tasks as benchmark.
%\end{itemize}
%
%To overcome these challenges and drive advancements in the field of multidimensional inversion in geophysics, we put forth three pivotal innovations:
%\begin{itemize}
%	\item[$\bullet$] Multi-dimensional Sample Generation.
%	\item[$\bullet$] Region-Growing Training Strategy.
%	\item[$\bullet$] Impedance Vectorization Projection.
%\end{itemize}
%
%These three pivotal innovations not only advance multidimensional inversion in geophysics but also have broader implications for the machine learning community. Our proposed innovations open doors for advancements in semi-supervised learning techniques across diverse domains. The discoveries and insights gained from this research can directly benefit and inspire developments in other semi-supervised regression tasks.
\begin{figure*}[htb]
	\centering
	\includegraphics[scale=0.8]{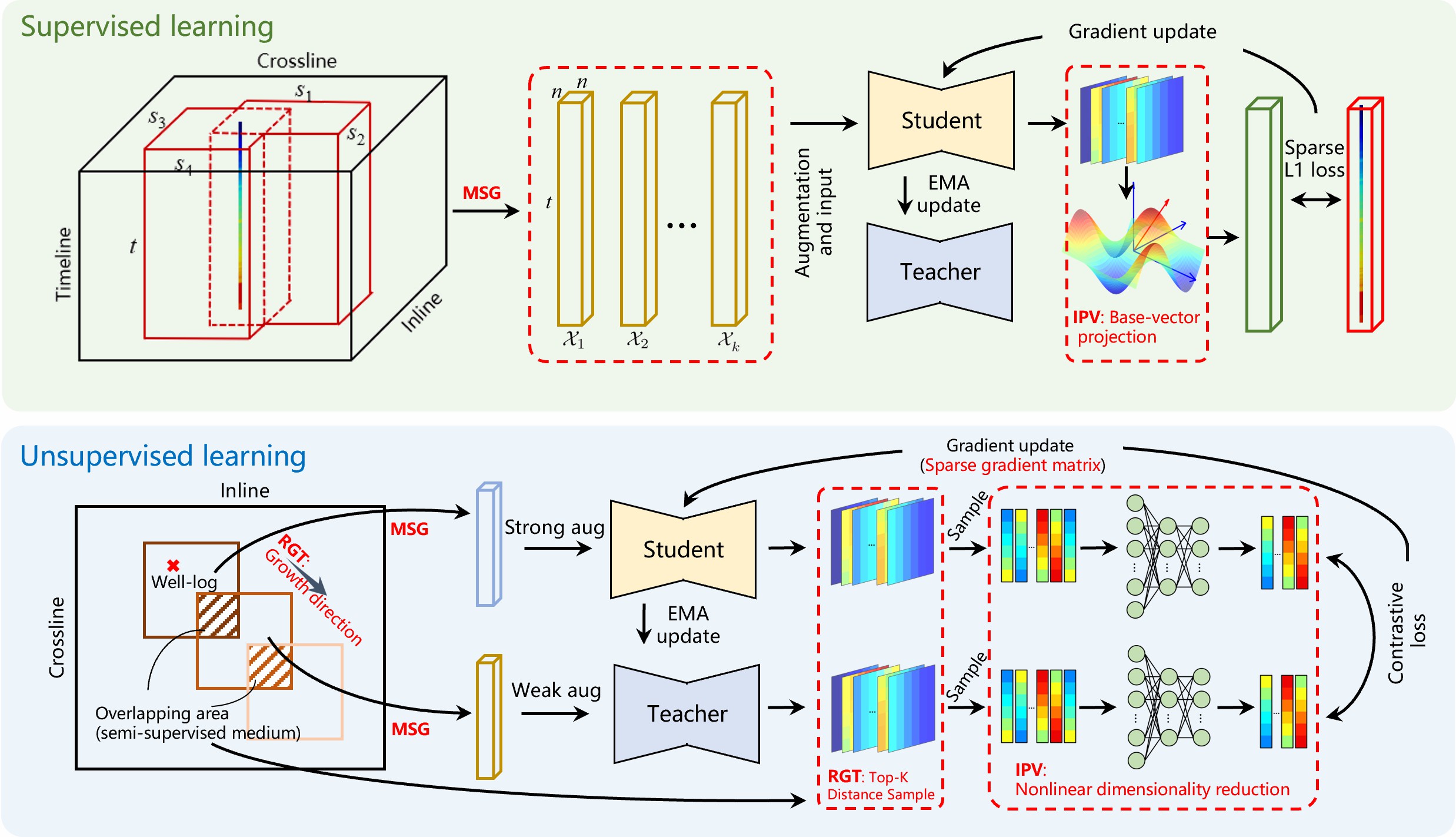}
	\centering\caption{The complete framework of ContrasInver, and the roles assumed by the individual innovations (red boxes or letters). In the supervised component, samples are generated by MSG, and the network outputs a vector that is projected onto basis vectors to obtain impedance (IVP). The supervised part is responsible for guiding and constraining the direction of the impedance vector. In the unsupervised component, RGT is utilized to gradually diffuse well log information and perform unsupervised learning based on IVP. One of the functions of IVP is to filter out certain abnormal components from the labels, preventing the issue of value confusion in semi-supervised learning where the regression task learns from the teacher network.}
	\label{framework}
\end{figure*}
\section{Approach}
\subsection{Semi-supervised inversion problem definition}
For a 3D seismic data $\textbf{D} \in \mathbb{R}^{t\times l_\text{i} \times l_\text{x}}$. All the seismic traces are represented as $\textbf{Z} = \{ {(z_i)}: i \in (1,...\mathcal{B})\}$, where $\mathcal{B} =  l_\text{i} \times l_\text{x}$. The labeled seismic traces are denoted as $\textbf{X}= \{ {(x_i, y_i)}: i \in (1,... \mu \mathcal{B})\}$, where $x_i$ represents the $i^\text{th}$ seismic trace and $y_i$ corresponds to the corresponding well log data, $y_i$ is a continuous value. 
Here, $\mu$ denotes the ratio of the number of labeled seismic traces $\mathcal{B}$ to the number of all seismic traces. In most open-source data, this ratio is less than 1e-4. 
%In the geophysical domain, a classical dataset like the F3 presents significant challenges due to the paucity of available well logs, with a ratio of $\mu$ being a mere 2.5e-5. This scarcity makes one-to-one training susceptible to overfitting, even during stages like pre-training, which typically have less demanding data requirements.
%Yet, seismic traces diverge from the norms of computer vision samples as they do not exist in isolation. 
%Rather, they are interconnected, bound by lateral associations and subject to the constraints of the seismic reflection axis. These properties allow for the potential to execute inversions even in the face of extremely sparse well logging circumstances. 
In geophysics, scarce datasets like the F3, with a well log ratio of $\mu$ at 2.5e-5, pose overfitting risks in training, even in typically less demanding pre-training stages. Seismic traces, however, unlike isolated computer vision samples, are interconnected, bound laterally and constrained by the seismic reflection axis, enabling inversions despite sparse well logging.
Concurrently, well logs are tied down by low-frequency limitations, as expressed by equation (\ref{lowfreq1}):
\begin{equation}
	y_i = y_i^\text{h} + t_i
	\label{lowfreq1}
\end{equation}
Where $y_i$ is the ground truth, $y_i^h$ signifies the well logging's high-frequency information relating solely to seismic reflection axis characteristics (texture), and $t_i$ indicates the depth-dependent low-frequency trend. Impedance values in two seismic data regions with identical characteristics but varying depths can differ due to the combined influence of feature and position constraints.
These unique aspects of well-seismic data necessitate the design of a semi-supervised inversion framework that is sensitive to and accommodating of these factors.

\subsection{Preliminary}
Viewed temporally, ContrasInver consists of two stages: pre-training and semi-supervised training. Spatially, ContrasInver encompasses two components: supervised and unsupervised training. During the pre-training phase, only supervised training is conducted. In the semi-supervised phase, both supervised and unsupervised training are executed concurrently.
Fig. \ref{framework} illustrates the supervised and unsupervised processes of ContrasInver. 

ContrasInver has some common components, as follows:

\textit{Mean Teacher: }
This is a common form of semi-supervised learning, where the student network is updated using gradients, and the teacher network is updated using Exponential Moving Average (EMA), equation (\ref{ema}).

\begin{equation}
	\mathcal{V}_t^\text{ema} =\lambda \mathcal{V}_{t-1}^\text{ema}+(1-\lambda) \mathcal{V}_t
	\label{ema}
\end{equation}
Where $\mathcal{V}_t^\text{ema}$ and $\mathcal{V}_{t-1}^\text{ema}$ are current and previous teacher parameters respectively, and $\mathcal{V}_t$ is the current student parameter.

\textit{Loss function: }
Supervised learning uses $\mathcal{L}_1$ loss, equation (\ref{regloss}).
\begin{equation}
\mathcal{L}_\text{Reg} =\sum_{d = 1}^{t} \sum_{h = 1}^{n} \sum_{w = 1}^{n} \parallel \hat{y}_{d,h,w}-y_{d,h,w} \parallel
		\label{regloss}
\end{equation}
Unsupervised (Contrastive) loss  uses cosine loss, equation (\ref{cosim}). 
\begin{equation}
	\mathcal{L}_\text{cosim}(v_1,v_2^\text{ema}) = \frac{v_1\cdot v_2^\text{ema}}{\text{max}(\parallel v_1 \parallel \cdot\parallel v_2^\text{ema} \parallel,\epsilon)} \label{cosim}
\end{equation}
 where $\epsilon$ is a small value to avoid dividing by $0$, $v_1$  and $v_2^\text{ema} $ are the feature vectors.
 
\textit{Data augmentation: }
It involves strong and weak augmentations. Weak ones include coordinate transformations like mirroring, rotation, and interpolation. Strong augmentation adds Gaussian noise (equation (\ref{nsa})) and gamma transformation (equation (\ref{gma})) specific to seismic data.
\begin{equation}
	\text{Aug}_\text{ns}(\mathcal{X}) = \mathcal{X} + \mathcal{G}, \ \mathcal{G} \thicksim \mathcal{N}(0,\sigma), \ \sigma \thicksim   \mathcal{U}(0.0,0.2) \label{nsa}
\end{equation}
\begin{equation}
	\text{Aug}_\text{gm}(\mathcal{X}) = \mathcal{X}^\gamma , \ \gamma \thicksim \mathcal{N}(1,0.1) \label{gma}
\end{equation}
The distributions of variables $\mathcal{G}$ and $\gamma$ are determined by the distribution of seismic data.

The following contents of this section are the innovations of this work.

\subsection{Multi-dimensional Sample Generation (MSG):}
MSG is a simple and practical method that was first proposed by us in this paper. It takes into consideration the characteristics of seismic data and effectively establishes the lateral continuity and vertical low-frequency constraints of seismic features. Additionally, it is easily scalable to higher or lower dimensions. MSG is not only suitable for semi-supervised learning but also applicable to supervised learning in densely well logs. It has the potential to become a sample construction paradigm for multi-dimensional inversion.

1D sample construction is simple, only requiring one-to-one regression learning on seismic traces at well log locations. 2D methods necessitate building planes that include well logs, with weights set to zero in areas without logs, typically needing at least five logs. While there aren't specific 3D inversion methods, other works can offer insight into 3D seismic data analysis. Given GPU limits, current 3D methods, like fault segmentation and seismic reconstruction\cite{dou2022md,dou2022mda}, divide seismic data into smaller $n\times n \times n$ blocks. Considering the imaging resolution of seismic, it is common to set $n=128$. As expressed in equation (\ref{lowfreq1}), although this approach can learn $y_i^\text{h}$, the $t_i$ are hidden in the well logs.

Though position embedding might seem promising, its application differs significantly in the context of ViT\cite{dosovitskiy2020image} or Transformers\cite{vaswani2017attention}. Unlike the absolute positions related to image patches or words, the depth of each seismic trace in seismic data correlates to geological time, rendering it relative. This results in a varying spatial resolution of depth across different seismic traces. Therefore, implementing position embedding would require assigning a unique embedding to each seismic trace, a process which is clearly impractical.

Sample construction is pivotal for specific tasks, shaping the learning process and framework design. In multi-dimensional inversion, it necessitates considering low-frequency constraints and maximizing data diversity in ultra-sparse scenarios, leading to the proposed MSG method.

When constructing sample cubes, a simple approach to address the low-frequency constraint issue is to only crop the crossline and inline axes while preserving the entire timeline. This preserves the complete depth position information, enabling the network to learn vertical correlations and constraints.

Let the original samples extracted by MSG from seismic data be denoted as $\mathcal{X}_k \in \mathbb{R}^{t \times s_\text{i} \times s_\text{x}}$. It is ensured that $\mathcal{X}_k$ contains at least one well log to facilitate supervised learning.
Suppose the well coordinates are $p_\text{i}^\text{raw}, p_\text{x}^\text{raw}$, and the height and width of the sample are $s_\text{i},s_\text{x}$, respectively, then the starting coordinates (upper left coordinates) of the sample are $p_\text{si}^\text{raw}, p_\text{sx}^\text{raw}$, and the range of  $p_\text{si}^\text{raw}, p_\text{sx}^\text{raw}$ is expressed as equation (\ref{startp}).

\begin{equation}
	\begin{aligned}
		p_\text{si}^\text{raw} &\thicksim \mathcal{U}(p_\text{i}^\text{raw}-s_\text{i}+\tau,p_\text{i}^\text{raw}+s_\text{i}-\tau)& \\
		p_\text{sx}^\text{raw} &\thicksim \mathcal{U}(p_\text{x}^\text{raw}-s_\text{x}+\tau,p_\text{x}^\text{raw}+s_\text{x}-\tau)&
	\end{aligned}\label{startp}
\end{equation}
where $\tau$ is the sample random bound set to ensure that subsequent semi-supervised training can have sufficient overlapping volume.

%We set the size of each $\mathcal{X}_k^\text{inp}$ to $t \times n \times n$, where $n$ is the length of inline and crossline after resize and $t$ is the timeline size of the original seismic data. To make sure that the interpolated seismic voxels are not severely distorted, we restrict the length of $\mathcal{X}_k$ inline and crossline i.e., $s_k \thicksim \mathcal{U}(\omega n,n / \omega)$, where $s_k$ is the original sampling inline or crossline length, $ \mathcal{S}$ is the matrix composed of $s_k$. 
%The parameter $\omega$ controls the range of amplification or reduction compared to the original data and is typically set to $0.5$.
Moreover, each cropped sample needs to be interpolated to the same size. 
Each $\mathcal{X}_k^\text{inp} \in \mathbb{R}^{t\times n \times n}$, with $n$ as the resized inline and crossline length, and $t$ as the original seismic data timeline size. To prevent severe distortion of interpolated seismic voxels,let $ s_\text{i}, s_\text{x} \thicksim \mathcal{U}(\omega n,n / \omega)$. The $\omega$ parameter, typically 0.5, controls the data amplification or reduction range.

Since the samples have been interpolated, the relative positions of the well logs also need to be adjusted accordingly. The adjustment of the interpolated well-log coordinates is expressed by the equation \ref{adjcoord}.

\begin{equation}
\begin{aligned}
p_\text{i} &= p_\text{i}^\text{raw} n/s_\text{i} \\
p_\text{x} &=  p_\text{x}^\text{raw} n/s_\text{x}
\end{aligned}\label{adjcoord}
\end{equation}
where $p_\text{i},p_\text{x}$ are the adjusted logging coordinates and $p_\text{i}^\text{raw},p_\text{x}^\text{raw}$ are the coordinates before sampling. The sample labels constructed using this method are incomplete (sparse), and in the dense regression, the weights of the regions without logs are set to 0.

The theoretical number of different samples that can be generated from each well log can be expressed by the equation (\ref{smp}).
\begin{equation}
\sum_{s_\text{i}= n \omega}^{n/ \omega} \sum_{s_\text{x}= n \omega}^{n/ \omega}s_\text{i} \cdot  s_\text{x}
\label{smp}
\end{equation}
Therefore, MSG greatly enriches the diversity of samples, laying a solid foundation for subsequent supervised and unsupervised learning tasks.

\subsection{Region-Growing Training Strategy (RGT)}

Fig. \ref{seismic-log} shows the notable lateral correlations in seismic data. Hence, we designed a semi-supervised proxy task using a region-growing strategy. After pretraining with MSG, closer locations to a well log predict impedance more accurately. The aim is to extend well log information throughout the seismic volume via iterative growth, using overlapping areas between inner and outer circles as the semi-supervised medium.

\subsubsection{RGT Growth Process} 
%The RGT growth process is performed using randomly generated virtual well locations.
%The virtual well location is obtained by doing a random offset to the real well location, the offset is gradually expanded with training, and the same sample crop approach is executed by the virtual well location so that the crop location is gradually diffused outward.
%The random offset is denoted by  $\alpha_\text{i},\alpha_\text{x}$, with  $\alpha_\text{i},\alpha_\text{x}$taking a range of values that grows linearly with the number of training steps, as defined in equation (\ref{alpha}).

The RGT growth process uses randomly generated virtual well locations. These are achieved by randomly offsetting real well locations, with the offset expanding with training, enabling gradual outward diffusion of the crop location. The random offset is denoted by $\alpha_\text{i},\alpha_\text{x}$, growing linearly with training steps, as defined in equation (\ref{alpha}).

\begin{equation}
	\begin{aligned}
		\alpha_\text{i} &\thicksim  \mathcal{U} (-\varphi \mathcal{I}_\text{cu}/\mathcal{I}_\text{all} \times  l_\text{i}/2, \varphi\mathcal{I}_\text{cu}/\mathcal{I}_\text{all} \times  l_\text{i}/2) \\
		\alpha_\text{x} &\thicksim  \mathcal{U} (-\varphi\mathcal{I}_\text{cu}/\mathcal{I}_\text{all} \times  l_\text{x}/2, \varphi\mathcal{I}_\text{cu}/\mathcal{I}_\text{all} \times  l_\text{x}/2)
	\end{aligned}\label{alpha}
\end{equation}
%where $\mathcal{I}_\text{all}$ is the total number of steps to be trained, $\mathcal{I}_\text{cu}$ is the current step, $l_\text{i}$ and $l_\text{x}$ are the inline and crossline lengths, respectively, and $\alpha_\text{i}$ and $\alpha_\text{x}$ are the range of offsets on the inline and crossline, respectively,
%$\varphi$ is a parameter controlling the termination of growth process, that is, when the training reaches $\mathcal{I}_\text{all} /\varphi $, the growth process is terminated, and then virtual wells are randomly set in the entire seismic volume and samples are generated. The coordinates of the virtual well $(p_\text{i}^\text{virt},	p_\text{x}^\text{virt})$ can be expressed as the equation (\ref{virtp}).

Where $\mathcal{I}_\text{all}$ is the total training steps, $\mathcal{I}_\text{cu}$ the current step, $l_\text{i}$ and $l_\text{x}$ are inline and crossline lengths, and $\alpha_\text{i}$ and $\alpha_\text{x}$ are the offset ranges on the inline and crossline. $\varphi$ controls the growth process termination. When training reaches $\mathcal{I}_\text{all} /\varphi$, growth ceases, virtual wells are randomly set throughout the seismic volume, and samples are generated. The virtual well coordinates $(p\text{i}^\text{virt}, p_\text{x}^\text{virt})$ are defined in equation (\ref{virtp}).

\begin{equation}
	\begin{aligned}
		p_\text{i}^\text{virt} &= p_\text{i}^\text{raw}+\alpha_\text{i} \\
		p_\text{x}^\text{virt} &= p_\text{x}^\text{raw}+\alpha_\text{x}
	\end{aligned}\label{virtp}
\end{equation}
Each well is randomly generated by two views (samples) in the manner described by MSG. This method of generating samples ensures that the twos have at least the overlapping volume of $(2\tau \omega)^2t$.

In RGT, the Euclidean distance between the center coordinates of two views and well coordinates is calculated. The greater distance, $\mathcal{X}_1$, undergoes strong augmentation to get $\mathcal{X}_1^\text{inp}$, and the shorter distance, $\mathcal{X}_2$, undergoes weak augmentation to get $\mathcal{X}_2^\text{inp}$.

$\mathcal{X}_1^\text{inp}$ is fed into the student and $\mathcal{X}_2^\text{inp}$ into the teacher to get $\mathcal{Y}_1$ and $\mathcal{Y}_2^\text{ema}$.

Including at least one sample with real wells per batch ensures the learning process stays on course. Hence, the final loss function for each batch can be expressed as equation (\ref{lidloss}).
\begin{equation}
	\mathcal{L}_\text{LID} = \eta_1 \mathcal{L}_\text{sup}+ \eta_2 \mathcal{L}_\text{uns}
	\label{lidloss}
\end{equation}
where $\eta_1$ and $\eta_2$ are weight parameters.

%The RGT needs to calculate the Euclidean distance between the center coordinates of two views and the well coordinates, and the farther distance is noted as  $\mathcal{X}_1$, which performs strong augmentation to get $\mathcal{X}_1^\text{inp}$, and the closer distance is noted as $\mathcal{X}_2$, which performs weak augmentation  to get $\mathcal{X}_2^\text{inp}$.

%Feed $\mathcal{X}_1^\text{inp}$ into backbone and $\mathcal{X}_2^\text{inp}$ into EMA backbone to get $\mathcal{Y}_1$ and $\mathcal{Y}_2^\text{ema}$.
%By employing the contrastive loss based on cosine similarity, the vectors in the overlapping region are encouraged to be more closely aligned. (\ref{cosim}).

%We include at least one sample containing real wells in each batch to ensure that the learning process does not deviate from the right track. Thus for each batch, the final loss function can be expressed as the equation (\ref{lidloss}).

\subsubsection{Distance TopK Sampling}\label{DTKS}

In the preceding section, we recognized the challenges in determining the correspondence between overlapping regions in augmented views, mainly due to coordinate transformations such as interpolation, mirroring, and rotation. Notably, interpolation can dramatically alter coordinates, thereby disrupting the one-to-one correspondence between the views.

While methods like PixPro have proposed a strategy for calculating the overlap area \cite{xie2021propagate}, it is only applicable to highly downsampled features (by 32x). When applied to images at the original resolution, it significantly consumes memory and slows down the training speed, making the training infeasible.

We propose Top-K distance sampling to alleviate this problem, which enables real-time calculation of overlapping regions between two cubes, $\mathcal{X}_1$ and $\mathcal{X}_2$, using just the CPU. Since there's no cropping or coordinate transformation augmentation along the timeline direction, we only compute the distance matrix of the two cubes in their original, unsampled forms, based on the coordinate matrices comprising the inline and crossline axes. These flattened matrices are referred to as $\mathcal{P}_1$ and $\mathcal{P}_2$, respectively. We calculate the distance matrix and sort it to obtain coordinates of the smallest $\mathcal{K}$ values, as shown in equation (\ref{tks}).
\begin{equation}
	\text{\textbf{F}}_\text{TKS}(\mathcal{P}_1,\mathcal{P}_2) = \text{arg}(\text{\textbf{F}}_\text{sort}(\textbf{\text{F}}_\text{dist} (\mathcal{P}_1, \mathcal{P}_2))_{1:\mathcal{K}}) = \textbf{c}_1,\textbf{c}_2 
	\label{tks}
\end{equation}
Where $\textbf{c}_1$ and $\textbf{c}_2$ are the sets of coordinates of the overlapping regions of $\mathcal{X}_1$ and $\mathcal{X}_2$, respectively, $|\textbf{c}_1|=|\textbf{c}_2|$, $\mathcal{K}\in [1,   (2 \tau \omega)^2 ] $.  The  $ \textbf{\text{F}}_\text{dist}(\cdot,\cdot)$ expression is as follows  equation(\ref{dista}).
\begin{equation}
	\textbf{\text{F}}_\text{dist} (\mathcal{P}_1,\mathcal{P}_2)  = ((\mathcal{P}_{1,\text{w}}-\mathcal{P}_{2,\text{w}})^2+(\mathcal{P}_{1,\text{h}}-\mathcal{P}_{2,\text{h}})^2)^{\frac{1}{2}}
	\label{dista}
\end{equation}
The loss calculation process can be expressed as equation (\ref{unsp}).
\begin{equation}
	\mathcal{A_\text{uns}} = \textbf{\text{F}}_\text{uns}( \textbf{\text{F}}_\text{smp}(\mathcal{F}_1,\textbf{c}_1), \textbf{\text{F}}_\text{smp}(\mathcal{F}_2,\textbf{c}_2)) \label{unsp}
\end{equation}
The function $\textbf{\text{F}}_\text{smp}(\cdot,\cdot)$ extracts the coordinate vectors of specific positions from the feature map based on a set of coordinates.
equation (\ref{tks}) is executed on the CPU, while equation (\ref{unsp}) is executed on the GPU. The computational complexity of equation (\ref{unsp}) is only $\mathcal{K} /n^2$ compared to methods like PixPro.

Top-K Distance Sampling efficiently computes only the necessary portions of overlapping regions, resulting in reduced computational demands without the need for GPU involvement. This enables the training process to accommodate larger batch sizes and increase training speed.

\subsection{Impedance Vectorization Projection (IVP)}
%In contrast to segmentation or classification tasks, one key difference in regression tasks is that the output is continuous and does not require label discretization. This poses a challenge in conventional semi-supervised learning approaches, as the student network may struggle to assess the quality of the labels, potentially leading to inferior performance.
%This subsection introduces IVP and its forward and backward propagation in both supervised and unsupervised processes. It also analyzes the potential factors that contribute to its effectiveness.

Unlike classification or segmentation tasks, regression tasks output continuous values and do not need label discretization. This can complicate traditional semi-supervised learning approaches, as it may be challenging for the student network to evaluate label quality, potentially undermining performance. This section introduces the IVP and discusses its application in both supervised and unsupervised processes, along with potential contributing factors to its effectiveness.

In the process of supervised learning, the network's output is no longer the final $\hat{y}$, but a vectorized representation $\hat{\mathbf{v}}$ obtained through projection. The relationship between the two can be expressed by the equation \ref{tr}.
\begin{equation}
	\hat{y}=\textbf{\text{F}}_\text{cosim}(\hat{\mathbf{v}},\mathbf{v}_\text{base})  \label{tr}
\end{equation}
Among them, $\mathbf{v}_\text{base}$  is a differentiable base vector, and its initial value is set to a vector of all ones, $\textbf{\text{F}}_\text{cosim}(\cdot)$ is cosine similarity.
The overall loss for vector $\hat{v}$ in supervised learning can be expressed as equation (\ref{vloss}).

\begin{equation}
	\begin{aligned}
		&\mathcal{L}_\text{sup} =\sum_{d = 1}^{t} \sum_{h = 1}^{n} \sum_{w = 1}^{n} \mathbbm{1}_{\{p\}} \parallel \textbf{\text{F}}_\text{cosim}(\hat{\mathbf{v}}_{d,h,w},\mathbf{v}_\text{base})-y_{d,h,w} \parallel \\
	\end{aligned}\label{vloss}
\end{equation}
The function $\mathbbm{1}_{\{p\}}$ is an indicator function where the value is 1 for locations corresponding to well log data and 0 for non-log locations. It is used to eliminate the influence of regions in the label where well log data is not available.

Although it is a composite loss function, its derivative form still remains as cosine loss. The $\mathcal{L}_1$ loss only controls the vector direction in equation (\ref{gd1}).
\begin{equation}
		\begin{aligned}
		\frac{\partial 	\mathcal{L}_{\text{sup}}(\hat{\mathbf{v}},\mathbf{v}_\text{base},y)}{\partial \hat{\mathbf{v}}}=
		\frac{\partial 	\mathcal{L}_{\text{sup}}(\hat{\mathbf{v}},\mathbf{v}_\text{base},y)}{ \partial \textbf{\text{F}}_\text{cosim}(\hat{\mathbf{v}},\mathbf{v}_\text{base})} \cdot
		\frac{\partial \textbf{\text{F}}_\text{cosim}(\hat{\mathbf{v}},\mathbf{v}_\text{base})}{\partial \hat{\mathbf{v}}}  \\
		=  \mathbbm{1}_{\{ \textbf{\text{F}}_\text{cosim}(\hat{\mathbf{v}},\mathbf{v}_\text{base})>y \}} 
		\cdot \frac{\partial \textbf{\text{F}}_\text{cosim}(\hat{\mathbf{v}},\mathbf{v}_\text{base})}{\partial \hat{\mathbf{v}}} 
		\end{aligned}
	\label{gd1}
\end{equation}
By following this approach, during the supervised process, the vectors representing impedance are not only supervised but also constrained using well logging data. Consequently, these constraints are subsequently transferred and applied to the unsupervised process, ensuring consistent guidance throughout the learning process.

In the unsupervised process, these vectorized impedances are non-linearly projected to a lower-dimensional space using an MLP. In this space, the distance between the student network and the teacher network is minimized. This process can be represented by the equation (\ref{supl}).
\begin{equation}
		\mathcal{L}_\text{uns} =  \textbf{\text{F}}_\text{cosim} (\text{MLP}_{\mathbf{W}_1,\mathbf{W}_2}(\hat{\mathbf{v}}),\mathbf{v}_\text{EMA})
		\label{supl}
\end{equation}
Where $\mathbf{W}_1$ and $\mathbf{W}_2$ are the differentiable parameters of the first and second layers of the MLP, and $\mathbf{v}_\text{EMA}$ represents the impedance vector of the teacher network after projection.

The MLP consists of two fully connected layers, a ReLU activation function, and a normalization layer. Let $\text{MLP}_{\mathbf{W}_1,\mathbf{W}_2}(\hat{\mathbf{v}}) = \hat{\mathbf{v}}_\text{p}$. The gradient expression for the vectorized impedance in equation (\ref{supl}) can be represented as equation (\ref{paruns}).
\begin{equation}
	\frac{\partial \mathcal{L}_\text{uns}(\mathbf{\hat{v}},\mathbf{v}_{\text{EMA}})}{\partial \mathbf{\hat{v}}} = 
	\frac{\partial \textbf{\text{F}}_\text{cosim} ( \hat{\mathbf{v}}_\text{p},\mathbf{v}_\text{EMA})}{\partial  \hat{\mathbf{v}}_\text{p}} \cdot \frac{\partial  \hat{\mathbf{v}}_\text{p}}{\partial \mathbf{\hat{v}}}
	\label{paruns}
\end{equation}
Where $\frac{\partial \textbf{\text{F}}\text{cosim} ( \hat{\mathbf{v}}\text{p},\mathbf{v}\text{EMA})}{\partial \hat{\mathbf{v}}\text{p}}$ represents the gradient due to the cosine similarity, and $\frac{\partial \hat{\mathbf{v}}_\text{p}}{\partial \mathbf{\hat{v}}}$ represents the gradient generated by the MLP, expanding it results in equation (\ref{weq}).

\begin{equation}
		\begin{aligned}
	\frac{\partial \hat{\mathbf{v}}_\text{p}}{\partial \mathbf{\hat{v}}} &= 
	\frac{\partial \hat{\mathbf{v}}_\text{p}}{\partial \text{\textbf{F}}_\text{ReLU}(\mathbf{h})} \cdot \frac{\partial \text{\textbf{F}}_\text{ReLU}(\mathbf{h})}{\partial \mathbf{h}} \cdot \frac{\partial \mathbf{h}}{\partial \hat{\mathbf{v}}} \\
	&= \mathbf{J}_{\mathbf{W}_2} (\mathbbm{1}_{\{\mathbf{h} > 0\}} \odot \mathbf{J}_{\mathbf{W}_{1}})
		\end{aligned}
	\label{weq}
\end{equation}
Where $\mathbf{h}$ is the hidden variable representing the output of the first layer of the perceptron, and the indicator function $\mathbbm{1}_{\{h > 0\}} $ is derived from the differentiation of the ReLU activation function. $\mathbf{J}_{\mathbf{W}_1}$ and $\mathbf{J}_{\mathbf{W}_2}$ are the Jacobian matrices with respect to $\mathbf{W}_2$ and $\mathbf{W}_2$, respectively.

By combining equations (\ref{gd1}), (\ref{paruns}) and (\ref{weq}), the gradient expression of the loss function with respect to the predicted impedance vector for a single batch is given by equation (\ref{paruns2}).
\begin{equation}
	\begin{aligned}
	\nabla  \mathbf{\hat{v}} =& 
	\eta_1 \cdot \mathbbm{1}_{\{ \textbf{\text{F}}_\text{cosim}(\hat{\mathbf{v}},\mathbf{v}_\text{base})>y \}} 
	\cdot \frac{\partial \textbf{\text{F}}_\text{cosim}(\hat{\mathbf{v}},\mathbf{v}_\text{base})}{\partial \hat{\mathbf{v}}} \\
	+& \eta_2 \cdot \mathbf{J}_{\mathbf{W}_2} (\mathbbm{1}_{\{\mathbf{h} > 0\}} \odot \mathbf{J}_{\mathbf{W}_{1}}) \cdot	\frac{\partial \textbf{\text{F}}_\text{cosim} ( \hat{\mathbf{v}}_\text{p},\mathbf{v}_\text{EMA})}{\partial  \hat{\mathbf{v}}_\text{p}}
	\end{aligned}
	\label{paruns2}
\end{equation}

We theorize that the Jacobian matrix  $ \mathbf{J}_{\mathbf{W}_2} (\mathbbm{1}_{\{\mathbf{h} > 0\}} \odot \mathbf{J}_{\mathbf{W}_{1}})$ in equation (\ref{paruns2}) plays a key role in the effectiveness of IVP. We hypothesize that during gradient propagation through the Jacobian matrix, outlier components get filtered out before they reach the impedance vector.

\begin{figure}[htb]
	\centering
	\includegraphics[scale=0.25]{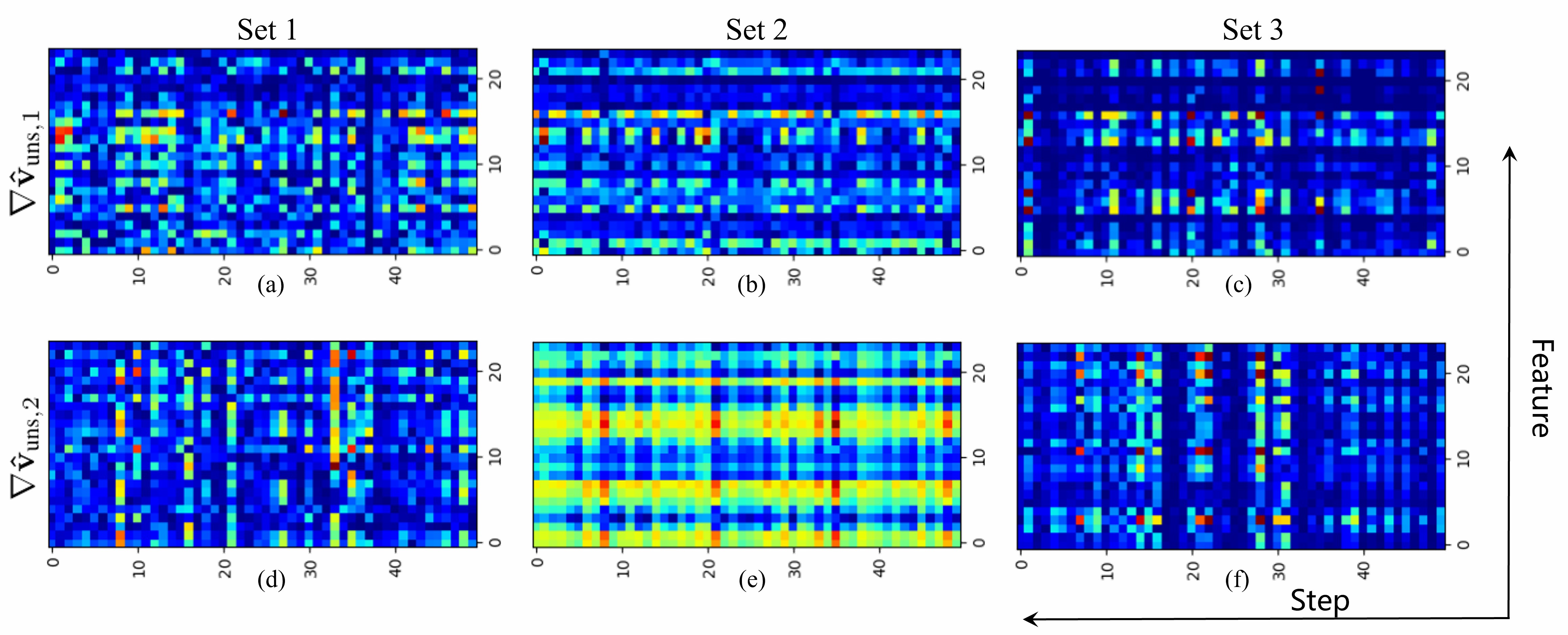}
	\centering\caption{Each subfigure has a horizontal axis representing  the number of steps. The vertical axis represents the feature dimensions of the impedance vector. All visualizations are performed using min-max normalization.}
	\label{set123}
\end{figure}

To verify this, we trained two models on the 16-well dataset from SEAM I, one using the full ContrasInver architecture and the other without the MLP component. The backbone network was HRNet, and the impedance vector length was set to 24. We conducted three tests: set 1, neither the student nor the teacher network was perturbed; set 2, the teacher network was perturbed; set 3, the student network was perturbed. Through these tests, we observed the effective mechanism of IVP.
Next, we disable parameter updates in the networks and continuously record the gradients $\nabla \mathbf{\hat{v}}_{\text{uns},1}$ and $\nabla \mathbf{\hat{v}}_{\text{uns},2}$ of both models with respect to $\mathbf{\hat{v}}_{\text{uns},1}, \mathbf{\hat{v}}_{\text{uns},2}$ for 50 steps. This process results in a visualized gradient matrix plot, which is referred to as Fig. \ref{set123}.

In Fig. \ref{set123}, set 2 is designed to observe the gradients backpropagated to the student network when there are a large number of outliers in the teacher network's outputs. In (b) and (e), the teacher network's outputs are perturbed by noise $\mathcal{G}_1$. In (e), we can observe significant fluctuations in the gradients of the loss with respect to the outputs, while in (b), under the same perturbation, the magnitude of the gradients is significantly lower than in (e). The backpropagated gradients in (b) are sparse, with many components filtered out. Set 3 aims to observe the gradients backpropagated when the student network's outputs deviate from the ground truth. We modify the input of the student network to be a normal distribution noise, resulting in (c) and (f). It can be observed that both (c) and (f) benefit significantly from the more accurate teacher network. This suggests that IVP does not blindly transmit sparse gradients in all cases. The stronger gradient intensity in (c) compared to (b) indicates that the IVP process considers the teacher network to be more accurate and thus retains more gradient components.

From this, we can draw the following conclusion: In IVP, a well-trained MLP adaptively selects the optimal subspace and applies the loss function, enabling the filtering of some incorrect gradient components transmitted to the impedance vector. This filtering process aims to retain as many valid components as possible, forming the basis for the effectiveness of IVP.

%One advantage of IVP is that both supervised and unsupervised learning objectives are based on cosine similarity. Although the supervised learning process applies L1 loss, it can be observed from the gradient expression in equation (\ref{gd1}) that it only affects the vector direction of the gradient and not the magnitude. In contrast, other contrastive semi-supervised multi-task training processes need to consider the changes in the weight distribution at the gradient intersection point between the supervised and unsupervised tasks\cite{zhong2021pixel,yang2022class,lee2022contrastive,alonso2021semi}. This allows us to set the weights for the two tasks ($\eta_1$ and $\eta_2$ in equation (\ref{lidloss})) and then no longer need to explicitly focus on them. 
%As indicated by equations (\ref{gd1}) and (\ref{paruns}), both the primary task (supervised) and the proxy task (unsupervised) are optimizing the impedance vectors, ensuring the consistency of the model optimization objectives. This allows the proxy task to effectively serve the primary task, enhancing the overall performance of the model.

IVP provides an elegant solution for semi-supervised learning in regression tasks. A cornerstone of IVP is its adeptness in adaptively filtering out the confusing information emanating from the teacher (EMA) network, thus preserving the crucial and effective feature for the learning process. This adaptive filtration is instrumental in enhancing the fidelity and robustness of the model. 
Furthermore, by vectorizing impedance and employing cosine similarity as a consistent optimization objective, IVP ensures a harmonious fusion of supervised and unsupervised learning phases. 
IVP is not only applicable to seismic inversion tasks but it a potential paradigm for regression-based semi-supervised learning tasks across various domains.

\section{Experiments}
\subsection{Experimental Settings}
\subsubsection{Synthetic data}
\begin{figure}[!h]
	\centering
	\includegraphics[scale=0.45]{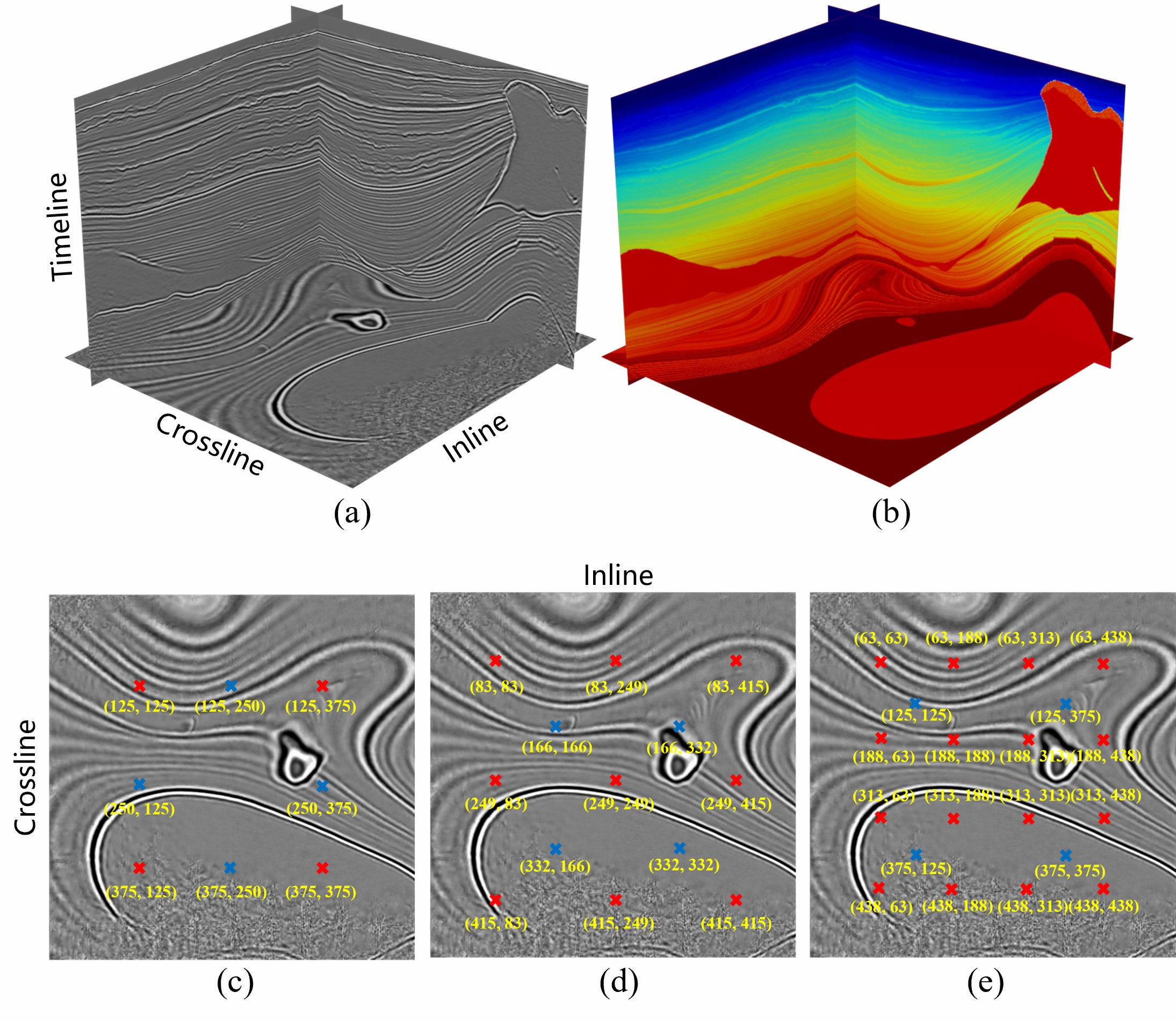}
	\centering\caption{(a) SEAM Phase I profile in grayscale. (b) GT. (c) Training and validation well locations with 4 logging wells. (d) Training and validation well locations with 9 logging wells. (e) Training and validation well locations with 16 logging wells. The red markings are for training logs and the blue are for validation logs.}
	\label{seamI}
\end{figure}

We employ the synthetic SEAM Phase I dataset in our experiments. Synthetic data offers the benefit of having complete 3D ground truth, enabling thorough validation of various methods. The data includes a complex salt body with significant impedance variations, both laterally and vertically \cite{wu2021deep}. Originally sized $600\times501\times502$ (timeline, inline, crossline), it's resized to $400\times501\times502$ for ease of training and inference. The denoised data is depicted in Fig \ref{seamI}.
For SEAM Phase I, conventional methods typically employ 30 or more logs \cite{wu2021deep,meng2021seismic}. In contrast, we have successfully reduced this number to 4, 9, and 16 logs, resulting in improved performance.

\subsubsection{Field data}
\textit{Netherlands F3:}
The F3 survey is a classic study widely used in various geophysical and imaging researches, including fault detection, salt body detection, seismic facies classification, seismic denoising, and seismic data reconstruction \cite{F3, dou2021attention, dou2022md, saad2022self, shafiq2016salsi, alaudah2019machine, qian2022ground, dou2022mda}. However, its application to machine learning-based impedance inversion is limited due to the availability of only four impedance logs in the original OpendTect project. Our ContrasInver method overcomes this limitation as it requires few logs to produce reasonable results. Fig. \ref{seismic-log} presents the F3 survey and the four impedance logs.

\begin{figure}[!h]
	\centering
	\includegraphics[scale=0.4]{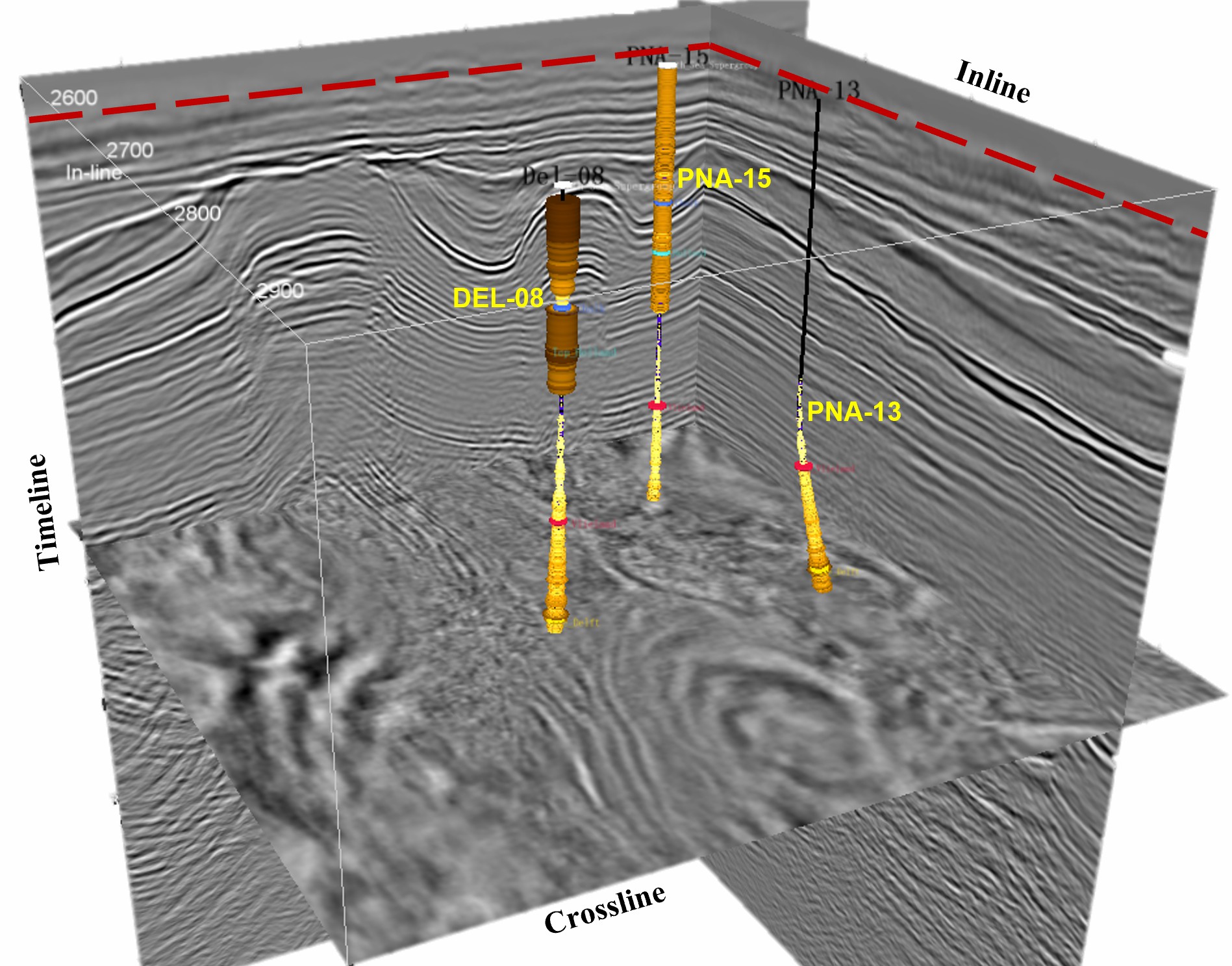}
	\centering\caption{The figure shows the Delft seismic data along with its three impedance well logs, all three of which are inclined wells. This is a typical example of extreme label less.}
	\label{exp6}
\end{figure}

\textit{Delft:}
The Delft survey, located in the West-Netherlands Basin (WNB) and provided by OpendTect \cite{DELFT}, offers only three well logs containing impedance as depicted in Fig. \ref{exp6} \cite{DINO}. Despite its limited data, this survey was successfully subjected to our initial data-driven inversion efforts.

\subsubsection{Comparison methods}
%There are widely used 1D CNN-based and semi-supervised methods for adversarial learning based on 1D CNN that are currently available, unfortunately, most of these methods are not open source, and we have implemented three representative works based on the literature \cite{das2019convolutional,wu2022seismic,wu2021semi}. 
%Among them Vishal's work \cite{das2019convolutional} is open source and we reimplemented it using Pytorch with reference to its code and papers, then we added residual structure to further improve the performance, similar methods are also compared in literature \cite{wu2021deep}. Since its structure is autoencoder, we name it ResNet-AE in the following. Based on this we refer to Wu's work adding attention structure to implement the literature \cite{wu2022seismic}, in the following it is named ResANet-AE.
%We also implemented a GAN-based semi-supervised method that references the work of Wu\cite{wu2021semi} and Wei\cite{hung2018adversarial}, the code uses Wei's and modifies it with reference to Wu's work. In the following it is named Semi-GAN. 
Despite the availability of various 1D CNN-based methods and semi-supervised adversarial learning techniques, most are not open-source. We implemented three notable models from the literature. Vishal's work \cite{das2019convolutional}, which is open-source, was reimplemented in PyTorch with the addition of a residual structure, referred to as ResNet-AE. We also used this model to implement Wu's method \cite{wu2022seismic}, incorporating an attention structure and naming it ResANet-AE. Additionally, a GAN-based semi-supervised method was implemented, named Semi-GAN, by referencing Wu \cite{wu2021semi} and Wei's \cite{hung2018adversarial} work.

%For the published multidimensional methods, currently, we only have access to Xinming's 2D work. However, we are unable to obtain their code, and considering its stringent data processing requirements and the need for low-frequency constraints, it is difficult to reproduce.Therefore, for the multi-dimensional approach, we combine MSG to modify Mean Teacher (MT)\cite{tarvainen2017mean} into a 3D inversion method (MT+MSG). In addition, MSG alone can also be used as a supervised multi-dimensional approach (MSG). By combining MT with the other two components proposed by us, two additional 3D methods can be created (MT+MSG+RGT, MT+MSG+IVP). This can be considered as a kind of ablation study. 

Regarding multidimensional methods, we currently only have access to Xinming's 2D work. Due to the lack of source code and the complexity of data processing requirements, it's difficult to reproduce. Thus, for multidimensional approaches, we modify Mean Teacher (MT)\cite{tarvainen2017mean} using MSG to create a 3D inversion method (MT+MSG). MSG alone can serve as a supervised multi-dimensional method (MSG). By combining MT with the two components we propose, we can develop two additional 3D methods (MT+MSG+RGT, MT+MSG+IVP), which can be viewed as an ablation study.

Due to the specificity of the inversion task, we were only able to implement a limited number of methods. This is in line with the reasons mentioned in the related work. The reason is that regression tasks cannot be easily assigned pseudo-labels, making pseudo-labeling methods difficult to reproduce. GANs require complete 3D labels, which are not available for sparse well logging data. Contrastive semi-supervised learning also relies on confidence scores for corresponding feature positions to identify high-quality features. Even if consistency regularization is applicable, it is still subject to various limitations. For example, many data augmentation techniques used in methods like UDA\cite{xie2020unsupervised} cannot be directly applied to seismic data, and methods like CCT\cite{ouali2020semi} rely on the assumption of semantic segmentation clustering. Overall, these methods are primarily developed to improve baseline models (e.g., MT) for semantic segmentation in natural images, and their innovations are not directly applicable to our task.

\subsubsection{Implementation Details}
The experiments are conducted on two 3090Ti GPUs. The batch size for 3D methods such as ContrasInver and MT is set to 6, while the batch size for parameter experiments of 1D methods is set to 256. The AdamW optimizer is used with a learning rate of 0.001.
In the semi-supervised approach, pre-training is performed for 5000 steps, followed by a semi-supervised process for 50000 steps. The non-semi-supervised method is trained for 55000 steps. 
Training ContrasInver for 55,000 iterations using PyTorch 2.0 with mixed precision and compile mode would take approximately 6 hours.

%In ContrasInver, the parameters are set as follows: $n=48$, $\tau = 4$, $\omega = 0.5$, and $\varphi=2$. The intuitive meaning of these parameters is as follows: $n=48$ indicates that the size of each sample is $t\times 48 \times48$, and we specify this size for all 3D methods in the comparative experiments. $\tau=4$ and $\omega = 0.5$ means that during the unsupervised process, randomly generated pairs of samples have at least $4\tau^2 \omega^2 t$ overlapping regions. $\varphi=2$ means that the growth process accounts for half of the semi-supervised steps. 
%This paper provides a parameter sensitivity experiment to demonstrate that these parameters are relatively optimal.

We set $n=48$, $\eta_2/\eta_1=10$, $\omega=0.5$, and $\varphi=2$. Among these parameters, $\omega=0.5$ and $\varphi=2$ have a mutual influence on seismic data types. We conduct parameter sensitivity study to demonstrate that the combination of $n$ and  $\eta_2/\eta_1$ we have chosen is relatively optimal.

Because no semi-supervised scheme for regression learning using MT directly has been found.
In this work, the semi-supervised process of the method without IPV utilizes the $\mathcal{L}_1$ loss.

\subsubsection{Evaluation metric}
In the synthetic data, we have access to the complete impedance ground truth, so we use two metrics: Mean Absolute Error (MAE) and Structural Similarity Index (SSIM). MAE represents voxel-level accuracy, and its expression is well-known and requires no additional parameter settings. Additionally, it is the only evaluation metric used during the validation process.
SSIM measures patch-level accuracy, and its expression is as equation (\ref{ssim}).
\begin{equation}
	\textbf{\text{F}}_\text{SSIM}(\textbf{I}_g', \textbf{I}_g) = \frac{(2\mu_{\textbf{I}_g'}\mu_{\textbf{I}_g}+c_1)(2\sigma_{\textbf{I}_g'\textbf{I}_g}+c_2)}{(\mu_{\textbf{I}_g'}^2+\mu_{\textbf{I}_g}^2+c_1)(\sigma_{\textbf{I}_g'}^2+\sigma_{\textbf{I}_g}^2+c_2)}
	\label{ssim}
\end{equation}
Where $\mu_{\textbf{I}_g'}$ is the mean of $\textbf{I}_g'$, $\mu_{\textbf{I}_g}$ is the mean of $\textbf{I}_g$,  $\mu_{\textbf{I}_g'}^2$ is the variance of  $\textbf{I}_g'$, $\mu_{\textbf{I}_g}^2$ is the variance of  $\textbf{I}_g$, $\sigma_{\textbf{I}_g'\textbf{I}_g}$ is the covariance of  $\textbf{I}_g'$ and $\textbf{I}_g$, $c_1$ and $c_2$ two variables to stabilize the division with weak denominator, see literature \cite{wang2004image} for details. 
The calculation of the equation (\ref{ssim}) is performed within a sliding window. We set the window size to be $7\times7\times7$.

Considering that complete ground truth data cannot be obtained from the field data, and thus SSIM cannot be calculated, we replace it with Mean Absolute Percentage Error (MAPE), which measures the percentage difference between predicted values and ground truth. Its expression is as equation (\ref{mape}).
\begin{equation}
	\textbf{\text{F}}_\text{MAPE} = \frac{1}{n}\sum_{i=1}^{n} \parallel \frac{y_i-\hat{y}_i}{y_i} \parallel \times 100\%
	\label{mape}
\end{equation}
Here, $y_i$ represents the true values, $\hat{y}_i$ represents the predicted values, and $n$ is the number of samples or instances in the dataset.

\begin{table}[!h]
	\centering
	\scriptsize
	\caption{Comparative experiments on synthetic data}
	\label{tab:ComparativeExperiments_syn}
	\begin{tabular}{@{}c|cc|cc|cc@{}}
		\toprule
		& \multicolumn{2}{c|}{4 well-log} & \multicolumn{2}{c|}{9 well-log} & \multicolumn{2}{c}{16 well-log} \\ \midrule
		& MAE     & SSIM    & MAE    & SSIM    & MAE     & SSIM    \\ \midrule
		ResNet-AE    & 0.3251 & 0.6133 & 0.2285 & 0.6949 & 0.1890  & 0.7550      \\
		ResANet-AE   & 0.3207 & 0.6218 & 0.1695 & 0.7802 & 0.1422  & 0.7838      \\
		Semi-GAN     & 0.3509 & 0.5956 & 0.1696 & 0.7320 & 0.1138  & 0.8522      \\
		MSG          & 0.1511 & 0.8865 & 0.1146 & 0.9041 & 0.0800  & 0.9375      \\
		MT+MSG       & 0.5651 & 0.6579 & 0.1456 & 0.8944 & 0.1159  & 0.9121      \\
		MT+MSG+RGT   & 0.5582 & 0.6725 & 0.1423 & 0.9079 & 0.1176  & 0.9111      \\
		MT+MSG+IVP   & 0.1325 & 0.9007 & 0.0559 & 0.9308 & 0.0208  & 0.9514      \\
		ContrasInver & \textbf{0.1105} & \textbf{0.9133} & \textbf{0.0269} & \textbf{0.9406} & \textbf{0.0153}  & \textbf{0.9589}      \\ \bottomrule
	\end{tabular}
\end{table}
\begin{table}[!h]
	\centering
	\scriptsize
	\caption{Comparative experiments on field data}
	\label{tab:ComparativeExperiments_fie}
	\begin{tabular}{@{}c|cc|cc@{}}
		\toprule
		& \multicolumn{2}{c|}{F3} & \multicolumn{2}{c}{Delft} \\ \midrule
		& MAE         & MAPE      & MAE          & MAPE       \\ \midrule
		ResNet-AE    & 1235     & 29.68\%   & -            & -          \\
		ResANet-AE   & 1149     & 28.13\%   & -            & -          \\
		Semi-GAN     & Nan         & Nan       & -            & -          \\
		MSG          & 467.3      & 10.70\%   & 1.353E+06     & 19.41\%    \\
		MT+MSG       & 523.8      & 12.54\%   & 1.874E+06     & 28.45\%          \\
		MT+MSG+RGT   & 519.7      & 12.10\%   & 1.738E+06     & 26.97\%          \\
		MT+MSG+IVP   & 296.1      & 6.447\%    & 7.726E+05     & 11.21\%          \\
		ContrasInver & \textbf{217.6}      & \textbf{5.080\%}    & \textbf{7.158E+05}     & \textbf{9.801\%} \\ \bottomrule
	\end{tabular}
\end{table}

\subsection{Experiments on synthetic data}
We divided the wells into test wells and validation wells in Fig. \ref{seamI}. The test set encompasses the entire seismic volume. 
Table \ref{tab:ComparativeExperiments_syn} presents the quantitative results on synthetic data, while Fig. \ref{exp1} illustrates the corresponding qualitative results.

\subsubsection{4 well logs} 
The extreme case of having only four well logs can better evaluate the performance of semi-supervised methods. The three 1D methods exhibit significant lateral discontinuity. Due to the limited availability of only four wells, there is a high risk of overfitting. Both ResNet-AE and ResANet-AE consistently reach their peak performance at 5000 training steps, after which the metrics on the validation set start to decline. The GAN-based method, on the other hand, reaches its peak at 8000 steps, but then the GAN starts to exhibit mode collapse.
Next is the multi-dimensional method. In general, the multi-dimensional method tends to reach the peak performance on the validation set around 30,000-40,000 steps. Using MSG as the baseline for the 3D method, ContrasInver shows significant improvements in both metrics (SSIM: +3.02\%, MAE: -26.9\%). MT+MSG+IVP also shows improvement (SSIM: +1.60\%, MAE: -12.3\%). On the other hand, the method without using IVP experiences a significant drop in performance, indicating that directly learning regression pseudo-labels without processing them can lead to severe performance loss. In Fig. \ref{exp1}, the qualitative results of (I-e) and (I-f) demonstrate a substantial loss of salt bodies, which can be attributed to the non-salt regions forcing the salt regions to learn their features in conventional regression semi-supervised learning. However, (I-g) and (I-h) successfully address this issue through IVP, highlighting the indispensability of IVP in semi-supervised learning for regression tasks.

\subsubsection{9 well logs} 
\begin{figure*}[!h]
	\centering
	\includegraphics[scale=0.4]{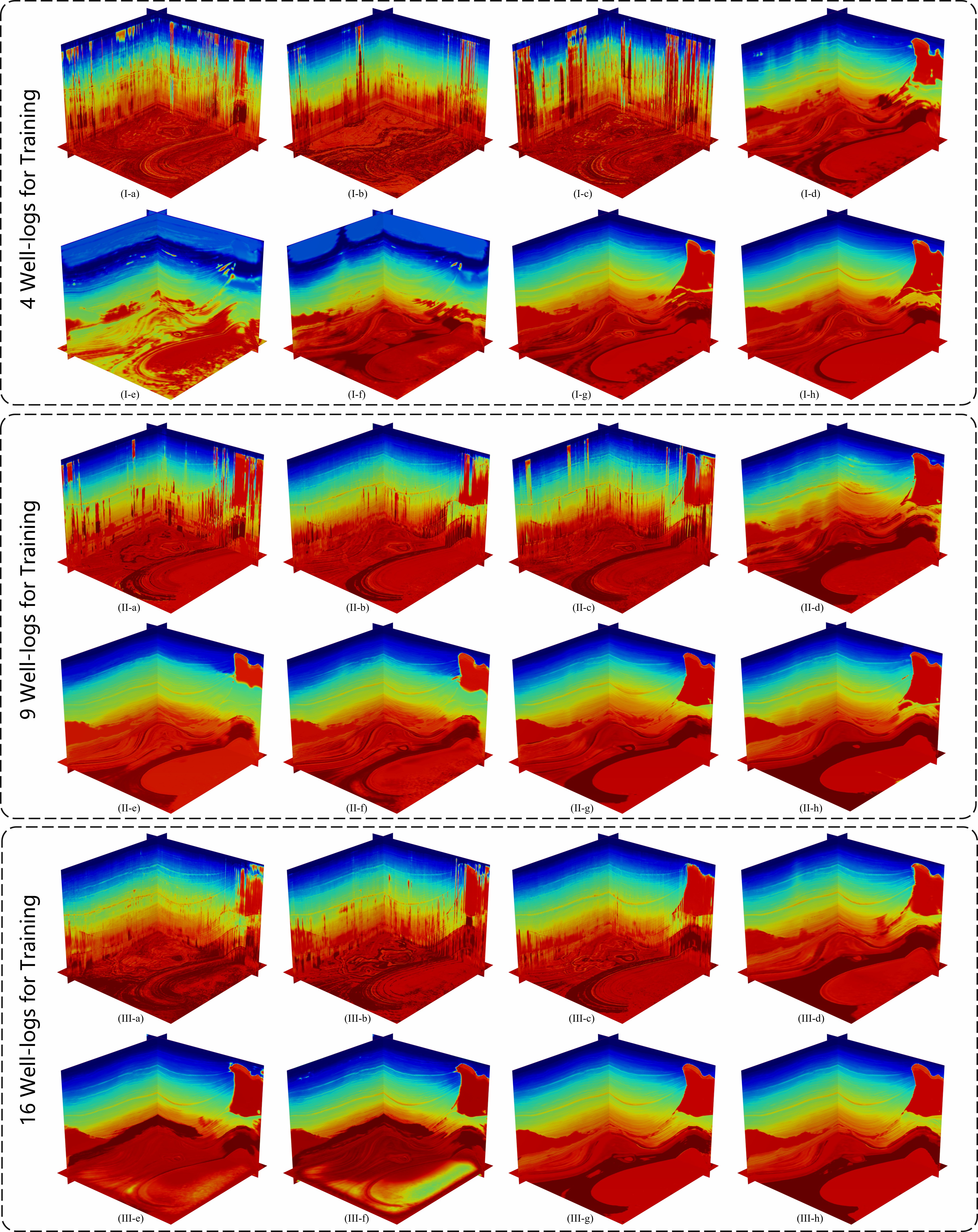}
	\centering\caption{Inversion results for synthetic data. I-$j$, II-$j$, and III-$j$ represent 4, 9, and 16 wells, respectively. $k$-a, b, c, d, e, f, g, and h correspond to ResNet-AE, ResANet-AE, Semi-GAN, MSG, MT+MSG, MT+MSG+RGT, MT+MSG+IVP, and ContrasInver, respectively. The GT corresponding to the synthetic data is Fig. \ref{seamI}.}
	\label{exp1}
\end{figure*}

The 1D methods still exhibit significant lateral discontinuity. In the multi-dimensional methods with MSG as the baseline, ContrasInver shows improvements of SSIM +4.03\% and MAE -76.5\%. MT+MSG+IPV shows improvements of SSIM +2.95\% and MAE -51.2\%.
Comparing (II-g) and (II-h) in Fig. \ref{exp1}, we can see that while they have similar inversion accuracy in the medium to low impedance regions, ContrasInver outperforms other methods significantly in describing high impedance regions. This is the benefit brought by RGT, which spreads the information from localized high-precision regions to the entire seismic volume. Instead of randomly generating samples throughout the seismic volume, which may lead to higher-precision samples learning from lower-precision samples, RGT helps maintain a more coherent learning process.

\subsubsection{16 well logs} 
In the 1D methods, Semi-GAN starts to demonstrate the potential of semi-supervised learning and appears to exhibit similar potential as the multi-dimensional methods. However, due to the inherent limitations of the 1D methods, they still struggle to maintain lateral consistency.
ContrasInver, trained on the 16-well dataset, is approaching the Ground Truth and shows significant improvements over the baseline model MSG, particularly in terms of voxel-level accuracy (MAE). The improvements relative to MSG are SSIM +2.2\% and MAE -80.9\%. MT+MSG+IPV shows improvements of SSIM +1.48\% and MAE -74.0\%.

\subsubsection{Conclusion from synthetic data} 
(1) Lateral discontinuity is an inherent limitation of 1D methods.
(2) Even without involving semi-supervised learning, MSG performs better than 1D methods.
(3) The semi-supervised framework can only work effectively when coupled with IVP.
(4) RGT offers a more reasonable training strategy and further enhances the performance beyond MT+MSG+IVP.

\subsection{Experiments on field data}
The main difference between field data and synthetic data lies in the increased diversity of noise. Field data not only contains additive noise but also includes coherent noise. Additionally, the accuracy of the labels (well logs) in field data is more ambiguous. The well data obtained in the early stages and the seismic data are not aligned, requiring manual well-seismic calibration. The accuracy of calibration depends on the precision of time-depth conversion and the expertise of the interpreter. In most cases, calibration can only be performed on a few clearly identifiable points, and interpolation is required between these points. This poses greater challenges for data-driven inversion. Table \ref{tab:ComparativeExperiments_fie}  presents the quantitative results, while Fig. \ref{exp2} and \ref{exp3} shows the qualitative results.

Given the  field data, which lacks a comprehensive ground truth, we are constrained in our evaluation methods. Quantitative metrics are assessed exclusively through well logs, while qualitative metrics are inherently dependent on empirical knowledge. Nevertheless, we can leverage our understanding of geological priors to develop a framework for qualitative assessment. We propose three guiding principles:
\begin{itemize}
	\item[$\bullet$] Significant lateral correlation: The inversion results should exhibit significant lateral correlation, where each seismic trace is correlated and continuous with its neighboring traces.
	\item[$\bullet$] Clear delineation of layers: The inversion results should be able to distinguish different geological layers, such as salt bodies, sandstones, etc. The boundaries of these regions should be clear or gradual.
	\item[$\bullet$] Clean inversion results: The inversion results should have minimal noise in different regions.
	\label{principles}
\end{itemize}

\subsubsection{Netherland F3}

The OpendTect original project did not provide impedance logs directly, and the impedance used for training was obtained from AI $=$ Vp $\times$ Density, where 'Vp' and 'Density' were provided by the original project.
For training, we limited the range of impedance to $[3500,6000]$. We chose a shorter log F02-1 as the validation well-log, and the other three participated in the training.

\begin{figure*}[!h]
	\centering
	\includegraphics[scale=0.5]{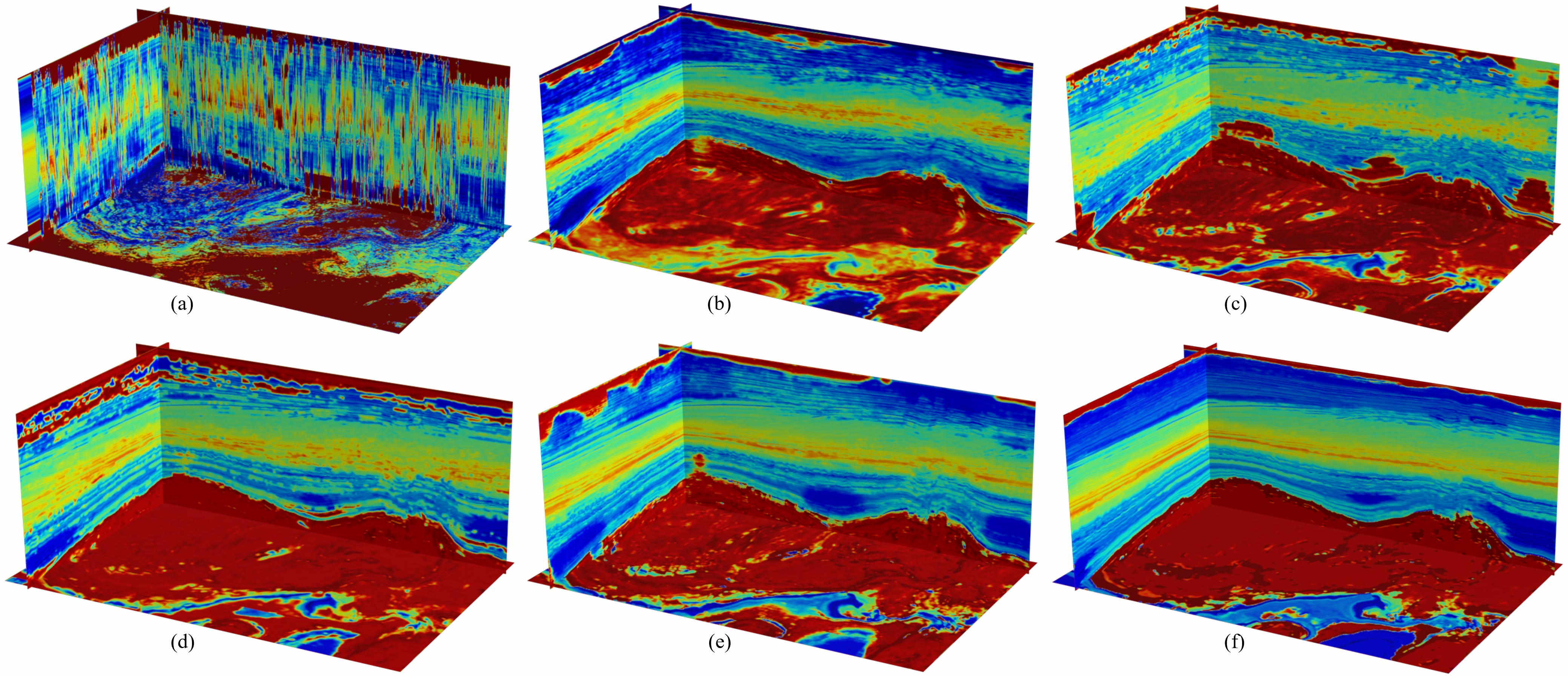}
	\centering\caption{Netherlands F3 using inversion results from wells F03-4, F03-2 and F06-1. (a) ResANet-AE, (b) MSG, (c) MT+MSG, (d) MT+MSG+RGT, (e) MT+MSG+IVP, (f) ContrasInver. Given the lack of comprehensive ground truth in field data, we offer a qualitative assessment approach based on a set of principles outlined in Checklist \ref{principles}.}
	\label{exp2}
\end{figure*}

In Table \ref{tab:ComparativeExperiments_fie}, the results of the 1D methods show relatively poor performance, and the 1D methods based on GAN fail to converge when faced with very few and low-quality labels. In Fig. \ref{exp2}, we only visualize the results of ResANet-AE, and it can be observed that it exhibits extremely chaotic results. Not only does it fail to maintain lateral continuity, but most of the predicted results also do not align with geological priors.

In terms of quantitative analysis, ContrasInver shows a significant improvement compared to the baseline model (MAE -53.4\%). However, when dealing with field data, it is important to focus more on qualitative results.
In Fig. \ref{exp2}, the (f) nicely divides F3 into four well-bounded impedance regions. Among them, ConstraInver accurately predicts the salt-body region with high impedance at the bottom. The middle and lower middle are sands, sand-stones, and claystones from Paleocene to Miocene, showing impedance values second only to the saltbody region.
By setting a threshold on impedance, it is possible to predict salt bodies with clear boundaries using only three wells. In comparison, some salt body segmentation methods require complete manual annotations to obtain accurate results \cite{saad2022self,zhang2022saltiscg}.
MT+MSG+IVP also exhibits a significant improvement in quantitative results (MAE -36.6\%). However, it is evident that its qualitative results are not as clear and boundary-defined as ContrasInver. ContrasInver demonstrates a better alignment with geological priors in terms of qualitative interpretation.

\subsubsection{Delft}
We used the 'AI final' provided in the original project as the impedance for training and validation, with a range of $[3.7 \text{e}6,1.5\text{e}7]$.
In Fig. \ref{exp6}, we chose a shorter log PNA-13 as the validation well-log, and the other two participated in the training.  
Because the two wells involved in training are deviated wells, it is not feasible to apply 1D methods. However, multi-dimensional methods are not constrained by this limitation.

\begin{figure*}[!h]
	\centering
	\includegraphics[scale=0.3]{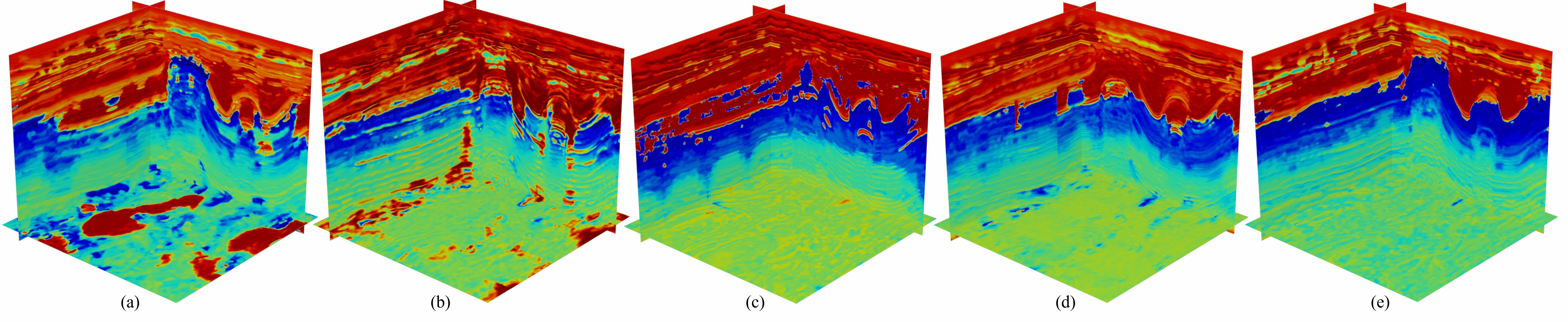}
	\centering\caption{DelftF3 using inversion results from two wells DEL-08 and PNA-15. (a) MSG, (b) MT+MSG, (c) MT+MSG+RGT, (d) MT+MSG+IVP, (e) ContrasInver.  Given the lack of comprehensive ground truth in field data, we offer a qualitative assessment approach based on a set of principles outlined in Checklist \ref{principles}.}
	\label{exp3}
\end{figure*}

In Fig. \ref{exp3}, ContrasInver demonstrates a more continuous result on the Delft dataset compared to other methods. The boundaries of the salt body region align perfectly with the corresponding seismic reflection axes. On the other hand, the result of MT+MSG+IVP shows the salt body crossing the boundaries, which may be attributed to the accumulation of errors during the semi-supervised learning process.
Both ContrasInver (MAE -47.1\%) and MT+MSG+IVP (MAE -42.9\%) show significant performance improvements. While the quantitative analysis does not reveal a substantial difference between the two methods, the qualitative interpretation highlights the superior performance of ContrasInver. It should be noted that this difference may be influenced by the use of a single short well for validation, potentially limiting the generalizability of the metric calculations.

\subsection{Parameter Sensitivity Study}
\begin{figure}[!h]
	\centering
	\includegraphics[scale=0.38]{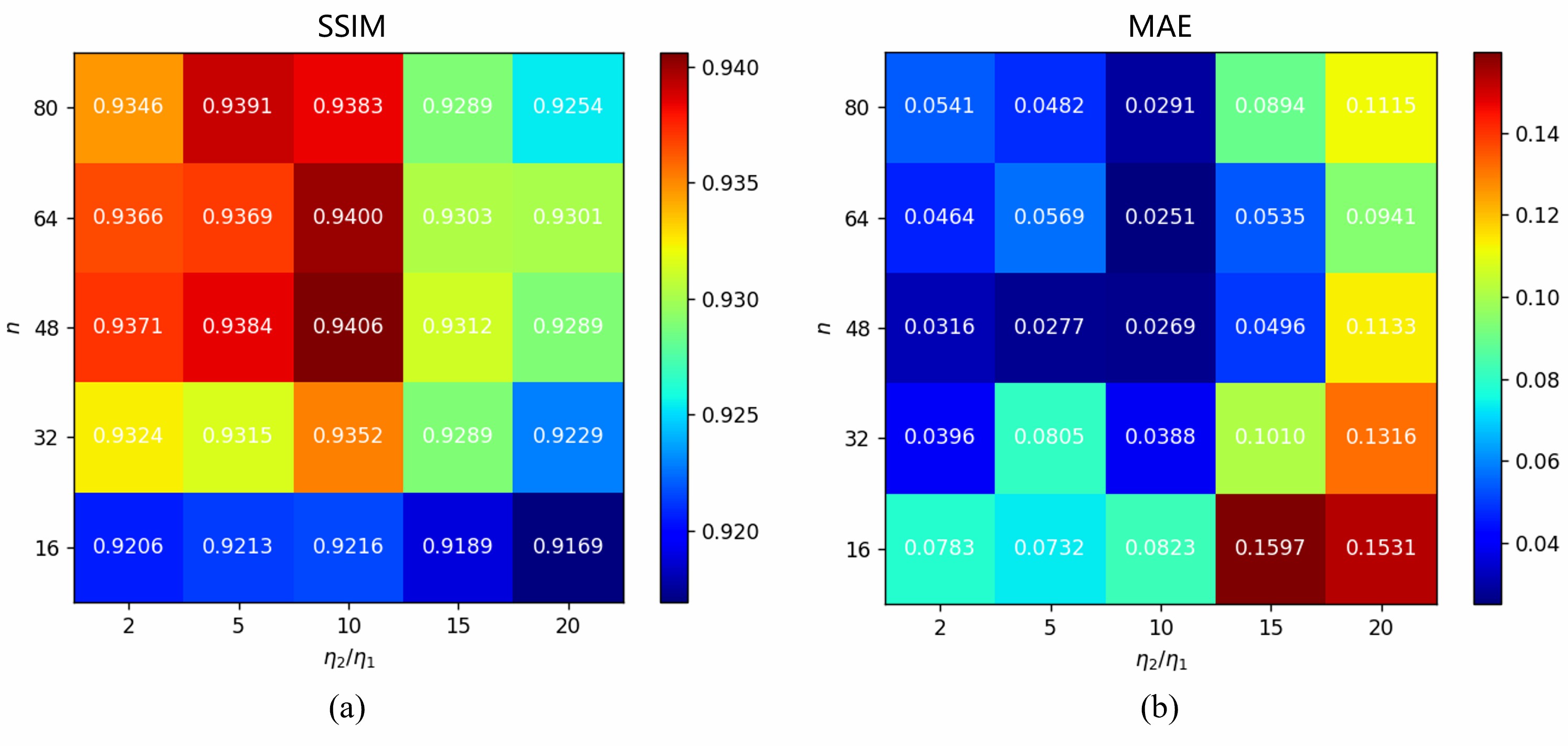}
	\centering\caption{The results of the cross-variable study for parameter sensitivity study. (a) SSIM metric, (b) MAE metric.}
	\label{exp7}
\end{figure}

The main hyperparameters we need to determine are $\eta_2/\eta_1$ and $n$. We can obtain an approximate range for  $\eta_2/\eta_1$ by calculating the ratio of gradients caused by supervised and unsupervised components in equation (\ref{paruns2}). However, manually computing this equation is not wise. By using the autograd tool in PyTorch, we found that the gradients caused by supervised learning are approximately $8-20$ times larger than those caused by unsupervised learning. We will consider the range $[2, 5, 10, 15, 20, 25]$ for  $\eta_2/\eta_1$ in our parameter sensitivity study. We set the range for $n$ as $[16, 32, 48, 64, 80]$,
due to GPU memory limitations, the batch sizes corresponding to these parameters are $[32, 16, 6, 4, 2]$. Training was conducted on SEAM I with 9 well logs.
We conducted a Cross-variable test and the results are shown in Fig. \ref{exp7}.

The optimal hyperparameters for the SSIM metric are $\eta_2/\eta_1=10$  and $n=48$, while for the MAE metric, the optimal combination is $\eta_2/\eta_1=10$ and $n=64$. The difference between these two optimal parameter combinations in terms of the SSIM and MAE evaluations is not significant. However, considering training speed, we choose the combination of  $\eta_2/\eta_1=10$  and $n=48$.


\section{Conclusion}
In this paper, we delve into the specific challenges encountered in multi-dimensional impedance inversion of seismic data using semi-supervised learning. To address these challenges, we introduce ContrasInver, which encompasses three pivotal innovations: MSG, IVP, and RGT. The MSG technique shows great promise as a paradigm for generating samples in multi-dimensional inversion. RGT creatively harnesses seismic lateral correlations to progressively propagate well log information. IVP tackles the critical issue of value confusion in semi-supervised regression tasks and has the potential to become a key component in such frameworks. Our experimental results demonstrate the remarkable superiority of our method, surpassing existing approaches in both qualitative and quantitative aspects on synthetic data. Furthermore, our method showcases groundbreaking advancements when applied to real-world field data.

\section*{Acknowledgment}
The authors are very indebted to the anonymous referees for their critical comments and suggestions for the improvement of this paper. Thanks to Xinming Wu for providing us with the data.

This is a pre-print version of the paper and we will do our best to answer all your questions before the paper is officially published (\textit{emindou3015@gmail.com}). The code for this paper will be public, and we will subsequently open source it here: https://github.com/douyimin/ContrasInver.

\bibliographystyle{IEEEtran}
\bibliography{references}

\end{document}